\documentclass{aa}

\usepackage{graphicx}
\usepackage{gensymb}
\usepackage[T1]{fontenc}
\usepackage{amssymb}
\usepackage{natbib}
\usepackage{tabularx}
\usepackage[utf8]{inputenc}
\usepackage[english]{babel}

\usepackage{txfonts}
\usepackage{ulem}
\usepackage{xcolor}

\usepackage{hyperref}
\renewcommand{\linenumbers}{}

\begin{document}

\title{WISE 12 $\mu{m}$ search for exozodi candidates within 10 parsecs }

\author{
Dong~Huang\inst{\ref{gzu}}
\and
Qiong~Liu\inst{\ref{gzu},\ref{cam}}
\and
Mark C. Wyatt\inst{\ref{cam}}
\and
Grant M. Kennedy\inst{\ref{war}}
}

\institute{
College of Physics, Guizhou University, Guiyang 550025, China\label{gzu}\\
\email{qliu1@gzu.edu.cn}
\and
Institute of Astronomy, University of Cambridge, Madingley Road, Cambridge CB3 0HA, UK\label{cam}\\
\email{wyatt@ast.cam.ac.uk}
\and
Department of Physics and Centre for Exoplanets and Habitability, University of Warwick, Gibbet Hill Road, Coventry CV4 7AL, UK\label{war}
}
\date{Received XXX; accepted YYY}

\abstract		
	{ The discovery of extra-terrestrial life is one of the ultimate goals for future exoplanet-seeking missions. One of of the major challenges these missions face is the possible presence of warm dust, referred to as ‘exozodiacal' dust, near the target stars or within their habitable zone. Therefore, it is critical to identify which stars possess exozodiacal dust and quantify their exozodiacal emission levels. 
	}	
	{ In this study, we conducted a search for exozodi candidates within 10 parsecs using the Reylé sample. We performed proper motion calculations and cross-matched the sample with the WISE and 2MASS database, resulting in 339 preliminary target samples. 
	}	
	{ We further analysed the infrared radiation characteristics of these targets, using spectral energy distribution (SED) fitting to predict photometric flux levels in the infrared and searching for $3\sigma$ excesses in the WISE W3 band. During the further selection process, we applied various analysis methods to perform rigorous validation.	
	}	
	{ We identified five exozodi candidates all of which are brown dwarfs (BDs). Given the clustering in candidate spectral types, we expect that these are not true exozodi candidates, rather the apparent excess arises from the inability of the BD photosphere models to accurately represent the SEDs of objects at the L-T transition. Indeed, for the object DENIS J025503.3-470049, excess is likely due to silicate clouds in the BD atmosphere. We suggest that a more stringent $5\sigma$ excess is required to infer excess for this spectral type.
	}	
	{ The detection rate (0/339) in our sample shows that less than 1\% of M stars have exozodi above 21\% excess levels. This is consistent with the rate of exozodi at similar level towards FGK stars in the Kennedy \& Wyatt sample (25/24,174). We provide upper limits on the 12 $\mu{m}$ exozodi emission for the sample, which is typically at 21\% relative to the star. For most stars, in particular the low mass M stars, this is the first such upper limit in the literature.
	} 


 \titlerunning{WISE 12 $\mu{m}$ search for exozodi candidates within 10 parsecs}
\authorrunning{Huang et al.}
\maketitle

\section{ Introduction}
\label{sec: Introduction}

As we enter a golden age of exoplanetary research \citep{2015JATIS...1a4003R,2013PASP..125..989A}, the scientific community is increasingly focused on achieving one of the ultimate goals in the field: the discovery of extra-terrestrial life. The ability to directly image habitable zone (HZ) exoplanets is crucial for this endeavour, as it allows for detailed analyses of their spectra to search for ‘biosignatures'—absorption features characteristic of molecules produced by biological activity, such as $H_{2}O$, $O_{2}$, $O_{3}$, $N_{2}O$, and $CH_{4}$ \citep{2018AsBio..18..739F,2018AsBio..18..663S}. Detection of such biosignatures would represent the first compelling evidence for the existence of life on other planets and concept missions such as Large Interferometer for Exoplanets (LIFE) \citep{2022A&A...664A..23K} aim to probe nearby stars in search of Earth-like exoplanets (exo-Earths) and to search them for these signals. The direct imaging of exo-Earths is currently limited to stars in the solar neighbourhood due to technological constraints and the faintness of the planetary signals compared to their host stars. High-resolution imaging instruments require nearby stars to ensure sufficient spatial resolution and contrast sensitivity, as these factors diminish significantly with increasing distance.

However, one of the major challenges these missions face is the possible presence of warm dust near the target stars or within their HZ. This dust, referred to as ‘exozodiacal' dust, is analogous to the ‘zodiacal' dust in our solar system, which originates from collisions and evaporation of comets and asteroids \citep{2010ApJ...713..816N,2010RAA....10..383K,2022MNRAS.510..834R}. Stars that harbour exozodiacal dust are said to have an ‘exozodi', and studies have shown that the flux from an exozodi can dominate that from an exo-Earth in the same planetary system, thus masking the presence of the latter \citep{2012PASP..124..799R}. Therefore, identifying analysewhich stars possess an exozodi and quantifying their exozodiacal emission levels is critical for future exoplanet-seeking missions.

The primary goal of our work is to search for more exozodi candidates or set constraints on the level of exozodi emission present towards a sample of stars likely to be searched for exo-Earths. The potential presence of warm dust orbiting a given star can be determined through photometry. The spectral energy distribution (SED) of a star describes its flux density emission at a given wavelength. As exozodis act as an additional source of flux density, the measured flux density for a star-exozodi system will be greater than is predicted by the lone theoretical stellar SED across the entire wavelength spectrum. Furthermore, this diﬀerence between the predicted and measured flux density will be greatest at the wavelengths where the exozodi emission is strongest. Assuming a single-temperature blackbody form of emission for exozodiacal dust \citep{2012PASP..124..799R}. Wien’s Law predicts that $300K$ dust has a peak emission wavelength of approximately 10 $\mu{m}$ . This means that for HZ dust, the ‘excess' emission will be greatest in the mid-infrared. The Wide-field Infrared Survey Explorer (WISE) \citep{2010AJ....140.1868W} operated across bands W1, W2, W3, and W4 with respective wavelengths of 3.4, 4.6, 12, and 24 $\mu {m}$; the W3 band is highly suitable for searching for HZ dust emission which best be detected as an excess in the W3 band. However, before a confident determination can be made of the presence of an exozodi, it is necessary to rule out false positive detections by consideration of photometric quality, stellar variability and contamination from other sources \citep{2018AJ....155..194E,2013A&A...555A.104A,2012MNRAS.426...91K}.

A separate goal of our work is to gain new insights into the conditions that correlate with the presence and emission intensity of warm HZ dust. Both the origin and evolution of warm dust is currently poorly understood; two diﬀerent mechanisms (‘in-situ' and ‘comet delivery') that provide explanations for these processes are explored in \cite{2013MNRAS.433.2334K}, and a ‘luminosity function' is used to determine the extent to which they can explain observed exozodis. This luminosity function is defined as the fraction of stars that possess an exozodi with a dust-to-star flux density ratio greater than a given level at 12 $\mu{m}$. Kennedy \& Wyatt analysed an initial sample of 24,174 solar-type stars observed by Hipparcos and identified 25 stars ($\sim0.1\%$ of the sample) exhibiting infrared excess attributed to exozodi, which we refer to as KW13 sample, hereafter. Producing a similar function for a diﬀerent sample of stars would thus provide information on the eﬀect of changing stellar properties on the frequency and strength of exozodi detections - which in turn might have implications for the applicability of theorized processes that govern the formation and development of exozodis \citep{2014A&A...570A.128E}.

The nearest stars are prime targets for exo-Earth imaging. The Reylé sample \citep{2021A&A...650A.201R} is a volume-limited sample of the nearby stars. Given the Initial Mass Function (IMF), most of these stars are lower in mass than those stars in KW13 sample. This means that we cannot know ahead of time what fraction will likely have detectable exozodi (i.e. from the KW13 statistics), and we may learn from this sample if exozodi are different around different stellar types.

This paper is arranged as follows: Section~\ref{sec:Sample selection} describes the sample selection process; In Sect.~\ref{sec:Identifying IR excesses}, the method used for identifying infrared excesses at 12 $\mu{m}$ is introduced; In Sect.~\ref{sec:results}, we verify the potential exozodi candidates; Sect.~\ref{sec:Discussion} provides an overview of the stellar properties of five exozodi candidates, concluding that these are not exozodi excesses. It also compares our exozodi detection rate with the rate from KW13 sample, and discusses the prospects for future exo-Earth search missions; Finally, Sect.~\ref{sec:Conclusion} summarises the conclusions of this work.

\section{Sample selection}
    \label{sec:Sample selection}

To analyse the objects in the Reyl\'{e} sample, we utilized data from AllWISE \citep{2014yCat.2328....0C}, Two Micron All Sky Survey (2MASS; \cite{2003yCat.2246....0C}), and Gaia Data Release 3 (DR3; \cite{2022yCat.1355....0G}). There are several different catalogues containing WISE data, and the AllWISE Source Catalog is the most complete catalog as it includes data from all of WISE's operational phases. Gaia DR3 provides fluxes in three optical bands ($G_{BP}$, $G$, and $G_{RP}$) and various astrophysical parameters derived from low-resolution BP/RP spectra. The 2MASS Point Source Catalog (PSC) provides near-infrared (NIR) fluxes in the $J$, $H$, and $K_{s}$ bands for point sources, corresponding to 1.2, 1.6, and 2.1 $\mu{m}$, respectively. 

\subsection{Proper motion calculation}

The Reyl\'{e} sample aims to achieve volume completeness based on current knowledge and provides a benchmark list of stars. It includes 562 stars, brown dwarfs (BDs), and exoplanets within 354 systems located within 10 parsecs of the Sun \citep{2022csss.confE.218R}. We focus on objects that could host exozodis (i.e. stars, white dwarfs, and BDs), since it would be hard to search for circumplanetary dust, as these are in close proximity to their stellar hosts. Therefore, it was necessary to cut the Reyl\'{e} sample to remove any exoplanets. The Reyl\'{e} catalog includes an ‘OBJ\_TYPE' column, which we used for this cutting. After removing 85 exoplanets, the sample was reduced from 562 objects to 477 objects. In our study, we refer to this as Sample A.

The proper motion of the various objects listed in the Reyl\'{e} sample is used to determine the co-ordinates of the Sample A objects. The Reyl\'{e} catalog provides proper motion values in RA and DEC for each entry, along with co-ordinates at a given epoch of observation, enabling us to determine object positions at different epochs. Since WISE operated in 2010, we can determine the epoch difference relative to 2010. We used Astropy to calculate the object's position at that time (i.e. its WISE co-ordinates). Additionally, the same method was applied to obtain its Gaia and 2MASS epoch co-ordinates.

\subsection{Cross-correlation}

In order to extract the WISE photometry data for the Sample A objects, it is necessary to cross-correlate the Reyl\'{e} catalog with the AllWISE Source Catalog. We use co-ordinates obtained through proper motion calculations and perform searches in VizieR\footnote{\url{https://vizier.cds.unistra.fr}}, an online astronomical catalog database that includes the AllWISE Source Catalog, via Astroquery. This is accomplished by providing the co-ordinates and specifying a search radius. The system returns data for any sources located within this search radius of the specified co-ordinates. For searches that return multiple objects, we identify the correct object by comparing the distance of each object to the specified co-ordinates—the closer object is considered to be the correct one. The W3 photometry obtained from this search will be compared with the predicted stellar flux to determine the presence of IR excess.

The Reyl\'{e} sample contains many binaries or multiple systems. Some of these sources share the same co-ordinates, while others have separations of less than 6\arcsec (i.e. slightly less than the angular resolution of any WISE band \citep{2010AJ....140.1868W}), both of which cannot be distinguished by WISE. For such stars we do not search for excess around each star individually, only in pairs or triads. There are 84 stars in 67 systems in total removed from the analysis on this basis, reducing the Sample A from 477 stars to 393.

We cross-correlated these 393 stars with ALLWISE using a search radius of 7\arcsec, which is slightly larger than the angular resolution of the W3 band, and obtained 382 matches. For the remaining 11 objects, we continued to enlarge the search radius until a WISE counterpart was matched for each object which corresponds to 17\arcsec. It was found that seven of these systems were matched to another star in the Reyl\'{e} sample, so they were considered binary systems. As a result, each was treated as one object, and these seven objects were excluded. For the remaining four objects that could not be identified with another star in the Reyl\'{e} sample, we checked their WISE images one by one and confirmed that for each object only one WISE source could be matched within their matching radius. Therefore, it is considered that these four sources are true WISE counterparts. 
In total, 386 ($382+4$) objects got WISE matches. However, since our goal is to examine the W3 fluxes from the sample, we subsequently excluded 27 objects due to poor W3 photometric quality (with the qph flag ‘U' or ‘X'). In total, 359 ($386-27$) objects got WISE matches with good quality W3 data.

\begin{figure}
        \centering
	\includegraphics[width=0.5\textwidth]{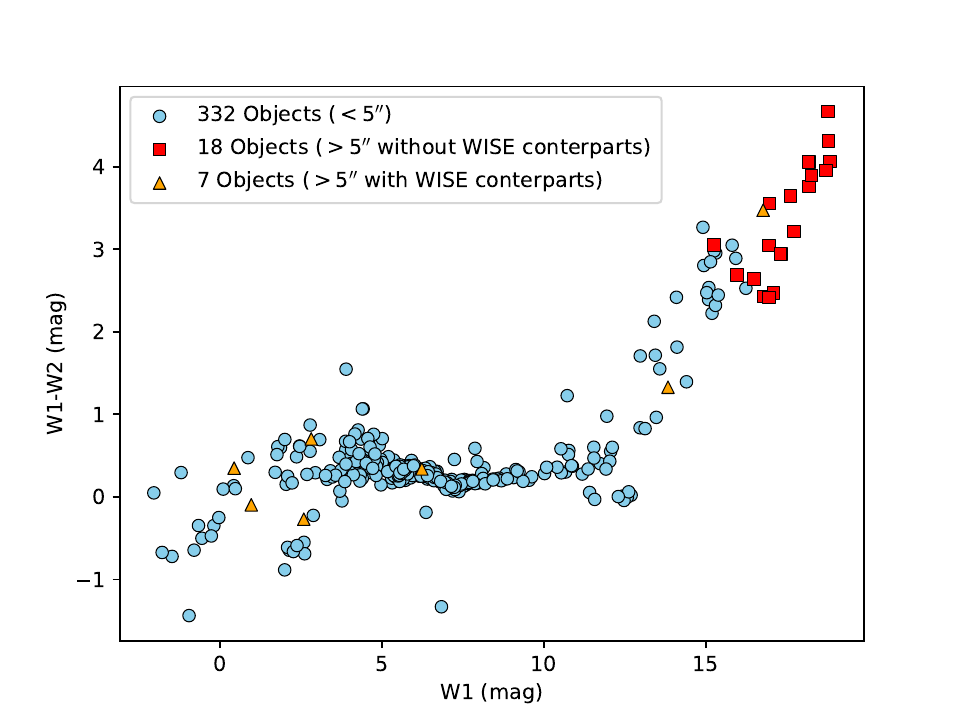}
	\caption{W1-W2 distribution with W1 about the 359 stars correlated with 2MASS. 
	The positions of the 359 stars cross-match with 2MASS on the $W1 - W2$ vs. $W1$ colour-magnitude diagram. The sky-blue circle marks the 332 matches within 5\arcsec, the orange triangle and red square represent the seven stars with WISE counterparts and 18 stars without WISE counterparts when enlarging the match radius. We note that there are two binary star systems matched to the same 2MASS counterpart, so they are treated as one object which cannot seen from this plot.}
	\label{fig:w1w2}
\end{figure}

Next, we cross-correlated these 359 objects with the 2MASS PSC using a search radius of 5\arcsec, resulting in 332 matches. For the remaining 27 objects, we continued to enlarge the search radius until a 2MASS counterpart was matched for each object. We then found that two binary star systems were matched to the same 2MASS counterpart, so they were treated as one object, and two objects were excluded. For the remaining 25 objects, we searched for the nearest WISE source to the 2MASS counterparts found with the enlarged search radius. We found that for seven objects this nearest WISE source was that which had already been matched with this object and so it is considered that these seven courses are true 2MASS counterparts. The remaining 18 objects were considered to have been matched with incorrect 2MASS counterparts, since their nearest WISE source was not that associated with the object in the Reyl\'{e} sample. Further inspection revealed that the WISE W1 magnitudes of these 18 objects were all fainter than 15, as Fig.~\ref{fig:w1w2} shows, making them too faint to be detected by 2MASS. We also cross-matched these 18 objects with other telescopes but did not obtain new data. Due to the lack of near-infrared data limiting the accuracy of photosphere predictions, we had to exclude these 18 objects. Therefore, a total of 339 ($332 + 7$) objects were matched with 2MASS data.

In summary, we obtained a final sample of 339 objects, which we refer to as Sample B. We cross-matched Sample B with Gaia data using a matching radius of 0.6\arcsec. As a result, we found that 312 objects had Gaia data, while 27 objects did not. For these 27 objects, their astrometry data from the literature is reported in \cite{2022csss.confE.218R} along with an estimate of the object's G magnitude G\_ESTIMATE. Given Gaia's magnitude limits at the bright ($G\cong2.5$ mag) and faint ($G\cong21$ mag) ends, this shows that among these 27 objects, 21 are too faint and six are too bright to be included in the Gaia catalog, explaining their absence. During the Gaia matching step, we did not exclude any objects. Table~\ref{tab:339 information} summarizes the stellar basic properties and photometric data for all stars in Sample B, arranged by distance, including spectral type, effective temperature, distance, and photometric data obtained from WISE W3. Note that here we only list the first ten objects for guidance regarding the tables form and content. The full version is available at the CDS.

\begin{table*}
	\centering
	\caption{Stellar basic properties and photometric data of the Sample B stars (portion).}
	\label{tab:339 information}
	\resizebox{180mm}{!}{
		\begin{tabular}{llccllccc} 
					\hline
			       Name        & SpT      &   D    &    $T_\mathrm{eff}$     &  $F_{12,obs}$     &    $F_{12,phot}$  &   $\chi_{12}$ &  ${F_{excess}/F_{12,phot}}$ & Upper limits \\

			                      &  & (pc) & (K) & ($mJy$) & ($mJy$) &  & &  \\
			                      (1)&(2)&(3)&(4)&(5)&(6)&(7)&(8)	& (9)\\
			\hline
Proxima Cen & M5.5 & 1.3020±0.0001 & 3000 & 912.631±12.585 & 922.238±44.335 & -0.156 & -0.010 & 0.201 \\
alf Cen A & G2 & 1.3459±0.0024 & 5750 & 122965.068±44334.671 & 168116.304±31509.799 & -0.826 & -0.269 & 0.976 \\
Barnard's Star & M3.5 & 1.8282±0.0001 & 3900 & 557.185±14.362 & 658.055±21.903 & -2.782 & -0.153 & 0.165 \\
Luhman 16 A & L7.5 & 1.9938±0.0003 & 1600 & 103.271±1.427 & 98.468±0.370 & 0.985 & 0.049 & 0.149 \\
WISEA J085510.74-071442.5 & Y2 & 2.2779±0.0124 & 550 & 1.090±0.127 & 1.009±0.017 & 0.596 & 0.081 & 0.645 \\
Wolf 359 & M6 & 2.4086±0.0004 & 3000 & 200.779±2.772 & 206.543±8.022 & -0.465 & -0.028 & 0.180 \\
HD 95735 & M1.5e & 2.5461±0.0002 & 3200 & 1906.919±26.322 & 1916.526±133.532 & -0.060 & -0.005 & 0.252 \\
alf CMa A & A1 & 2.6371±0.0110 & 11750 & 77813.832±23920.548 & 103271.444±12680.773 & -0.933 & -0.247 & 0.793 \\
G 272-61 A & M5 & 2.7195±0.0055 & 3000 & 388.108±5.380 & 396.754±12.441 & -0.391 & -0.022 & 0.167 \\
Ross 154 & M3.5e & 2.9760±0.0003 & 3400 & 336.713±4.650 & 343.918±27.955 & -0.224 & -0.021 & 0.280 \\
		\hline
		\end{tabular}
	}
Notes. Column (1): Name of Sample B objects. Col. (2): Spectral type quoted from the Reyl\'{e} sample.
 Col. (3): Stellar distance from Earth quoted from the Reyl\'{e} sample. Col. (4): Effective temperature obtained from VOSA SED fitting. Col. (5): Observed data in the W3 band from ALLWISE. Col. (6): Predicted photospheric data in the W3 band obtained from VOSA SED fitting. Col. (7): Flux excess significance, $\chi_{12}$, in the W3 band. Col. (8): ${F_{excess}/F_{12,phot}}$ at 12 $\mu{m}$. Col. (9): Upper limit of $F_{excess}/F_{12,phot}$ at 12 $\mu{m}$, taken as the larger value between $3\sigma_{tot}/F_{12,phot}$ and $3\sigma$.
(A portion of the table is shown here for guidance regarding its form and content. The full version is available at the CDS.)

\end{table*}

\section{Identifying IR excesses}
\label{sec:Identifying IR excesses}

The next step is to predict the stellar photosphere emission and compare it with the observed emission to identify whether an IR excess is present \citep{2006ApJ...636.1098B}. 
The method involves SEDs to the available photometry, known as ‘SED fitting'. 
Available photometric measurements are used to fit the stellar atmosphere and predict the photospheric flux at longer wavelengths, which is then compared to the WISE W3 observations.

\subsection{VOSA SED fitting}

We collected photometric data from optical and IR bands, summarized in Table~\ref{tab:339 information}, and used VO SED Analyzer (VOSA) \footnote{\url{http://svo.cab.inta-csic.es/theory/vosa/}} \citep{2008A&A...492..277B} to analyse the SEDs of each object in Sample B. VOSA, a Virtual Observatory tool, can automatically build SEDs for thousands of objects across a wide range from ultraviolet to IR, utilizing numerous photometric catalogues. VOSA compares the photometric data with different theoretical model sets and applies various statistical methods to determine which model best reproduces the observed data. 
The physical parameters of each object, such as effective temperature and luminosity, are then estimated from the best-fitting model.

VOSA applies the extinction law of \cite{1999PASP..111...63F}, incorporating the IR improvements by \cite{2005ApJ...619..931I}, to deredden the SEDs. The goodness of fit for SED fitting is assessed using the $vgfb$ parameter. Internally, VOSA calculates $vgfb$ by enforcing $\sigma(F_{obs}) > 0.1 * F_{obs}$, where $\sigma(F_{obs})$ is the error in the observed flux ($F_{obs}$). This parameter is a pseudo-reduced $\chi^2$ used within VOSA. It is particularly useful for avoiding the over-weighting of photometric points with underestimated flux errors. A threshold of $vgfb < 15$ is used to indicate a good fit \citep{2023AN....34430116S}.

In our Sample B, the objects can be broadly classified into five sub-samples basing on their spectral types from the Reyl\'{e} sample: D (white dwarfs), AFGK, M, L, and TY. For the D (white dwarfs) and AFGK sub-samples, we used photometric data from Gaia, 2MASS, and WISE to perform the SED fitting \citep{2024AJ....168..157C}. For the M, L, and TY sub-samples, we used photometric data from 2MASS and WISE to perform the SED fitting as in \citet{2017MNRAS.469..579B,2024ApJ...960..105Z}. Furthermore, for each of these five sub-samples, we selected the corresponding theoretical stellar atmosphere models available in VOSA to perform the SED fitting process:

\textendash~ For the nine white dwarfs, we used the Koester WD models \citep{2010MmSAI..81..921K,2009ApJ...696.1755T} to perform the SED fitting.

\textendash~ For the 55 AFGK-type stars, we fitted their SEDs using the Castelli-Kurucz model \citep{2003IAUS..210P.A20C}.

\textendash~ For the 226 M-type stars, which belong to the biggest subsample, we performed the SED fitting using the BT-Settl model atmospheres \citep{2006MNRAS.368.1087B,2009ARA&A..47..481A,2013MSAIS..24..128A,2012RSPTA.370.2765A,2011ASPC..448...91A,2007A&A...474L..21A,2003IAUS..211..325A} for stars with spectral type earlier than M7 (effective temperatures $T_\mathrm{eff}>2700$ K) and the AMES-DUSTY models \citep{1997JChPh.106.4618P,1993A&A...271..587G,2000ApJ...542..464C,2001ApJ...556..357A} for those with spectral type later than M7 ($T_\mathrm{eff}\leq2700$ K) \citep{2017MNRAS.469..579B}.

\textendash~ For the 20 L-type stars, we analysed their SEDs using the BT-DUSTY \citep{2011SoPh..268..255C}, BT-Settl (AGSS2009) \citep{2006MNRAS.368.1087B}, and DRIFT-PHOENIX \cite{2003A&A...399..297W,2011A&A...529A..44W,2009A&A...506.1367W,2008ApJ...675L.105H,2008A&A...485..547H,2006A&A...455..325H,1999JCoAM.109...41H,2003IAUS..210...19B,2004A&A...414..335W} models.

\textendash~ For the 29 TY-type stars, we used six theoretical models to perform the SED fitting: ATMO 2020 \cite{2020A&A...637A..38P}, BT-Settl-AGSS2009, AMES-Cond 2000, BT-COND, Saumon 2012 \citep{2012ApJ...750...74S}, and Morley 2012 \citep{2012ApJ...756..172M}.

\subsection{Finding W3 excesses}

After the VOSA fitting was completed, we extracted the predicted photospheric W3 flux density $F_{12,phot}$ and its corresponding uncertainty $\sigma_{phot}$ for each object. We then compared these with the measured W3 flux density as observational flux $F_{12,obs}$, by converting the measured W3 magnitudes to flux density.

Objects are deemed to have a significant excess of flux in W3 (i.e. regarded as potential exozodi candidates) based on the criterion: the flux excess significance at 12 $\mu{m}$, $\chi_{12}$. This significance is defined as the difference between the observed and predicted photospheric flux, that is the excess flux $F_{excess} = F_{12,obs} - F_{12,phot}$ divided by the total error, which must be greater than 3, that is

\begin{equation}
	\chi_{12} = \frac{F_{excess}}{\sigma_{tot}}\geqslant3~,
	\label{equation:1}
\end{equation}

\noindent where

\begin{equation}
	\sigma_{tot}=\sqrt{{\sigma_{obs}^2}+{\sigma_{cal}^2}+{\sigma_{phot}^2}}~.
	\label{equation:2}
\end{equation}

The total uncertainty, $\sigma_{tot}$, is composed of three uncertainties: observation uncertainty $\sigma_{obs}$, calibration uncertainty $\sigma_{cal}$, and photosphere model uncertainty $\sigma_{phot}$. The $\sigma_{obs}$ is obtained from the observational uncertainty provided for the W3 band. The absolute calibration uncertainty of WISE is estimated to be 4.5\% for the W3 band \citep{2011ApJ...735..112J}. So the calibration uncertainty of W3 should be $\sigma_{cal} = 4.5\% * F_{12,obs} $. The absolute model uncertainties are provided by VOSA as $\Delta{F_{tot}}/F_{tot}$, where $F_{tot}$ represents the total flux of the star, and $\Delta{F_{tot}}$ represents the error in the total flux of the star. So the model uncertainty of W3 should be $\sigma_{phot} = \Delta{F_{tot}}/F_{tot} * F_{12,phot}$. There are a total of ten stars that meet this criterion (hereafter referred to as criterion A, where $A\equiv\chi_{12}>3$).

\begin{figure}
	\includegraphics[width=0.49\textwidth]{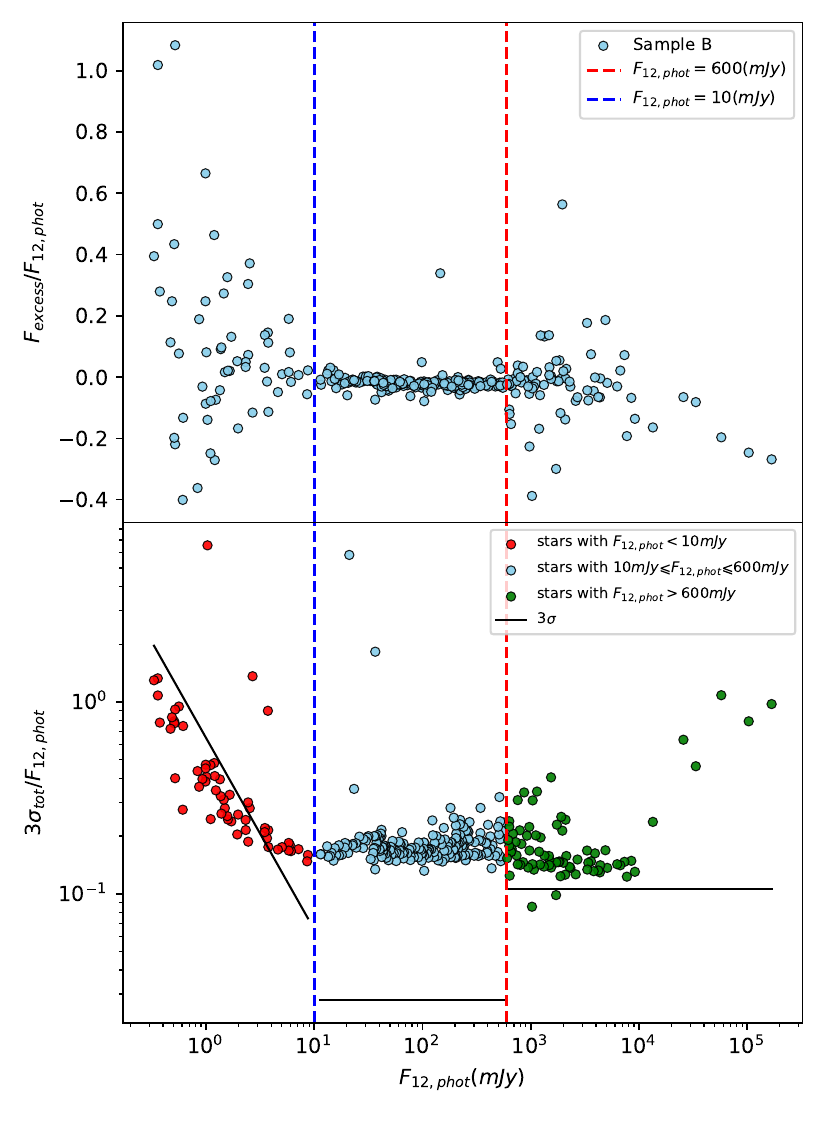}
	\caption{Scatter plot of $F_{excess}/F_{12,phot}$ versus $F_{12,phot}$ for Sample B stars. The vertical dashed lines represent the boundaries at $F_{12,phot} = 10 mJy$ (blue) and $F_{12,phot} = 600 mJy$ (red), indicating different flux regimes. The bottom panel is divided into three regions based on $F_{12,phot}$: red dots represent stars with $F_{12,phot} < 10 mJy$; blue dots represent stars with $10 mJy \leqslant F_{12,phot} \leqslant 600 mJy$; green dots represent stars with $F_{12,phot} > 600 mJy$. And the black line represents the empirical measure of uncertainty $3\sigma$ for each regime.}
	\label{fig:figure2.pdf}
\end{figure}

Given that for most stars the observed flux is consistent with purely photospheric emission, the distribution of observed fluxes about the predicted photospheric level is an empirical measure of the true uncertainty. Therefore, the uncertainty on the excess measurements can be assessed by looking at the distribution of $F_{excess}/F_{12,phot}$. This is shown in Fig.~\ref{fig:figure2.pdf} where it is clear that there are three regimes. For all stars the average is close to zero, which is because these observations detect photospheres for most stars. The distribution about the mean is an empirical measure of the uncertainty in the excess, which has a narrow distribution for intermediate brightnesses, much higher for bright stars (due to saturation), and increasing to fainter stars in the <10 mJy regime due to approaching the sensitivity limit of WISE. 
We fit the distribution of either the flux density ratio $F_{excess}/F_{12,phot}$ (or the flux density $F_{excess}$) to a Gaussian with mean $\mu$ and standard deviation $\sigma$. Then we identify stars that deviate from this distribution by $B=(F_{excess}/F_{12,phot} -\mu)/\sigma >3$ (or $B=(F_{excess} -\mu_{excess})/\sigma_{excess} >3$) as excess candidates. We refer to this criterion as Criterion B, hereafter. We determine an empirical measure of the uncertainty $\sigma$ in each regime in a different way.

\begin{figure}
	\includegraphics[width=0.5\textwidth]{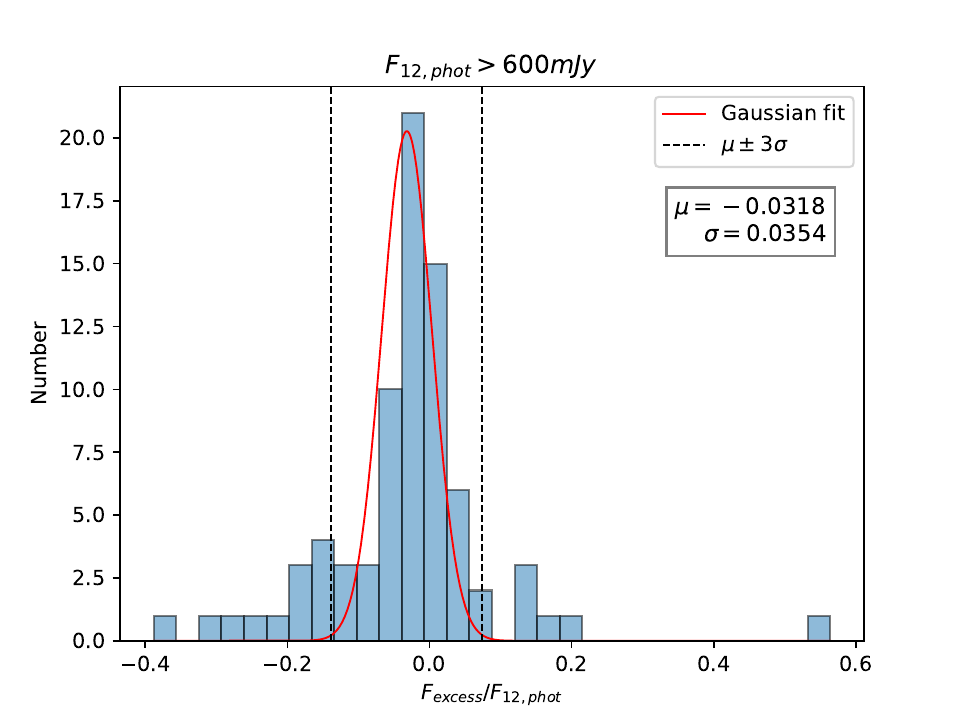}
	\caption{Histogram of the flux density ratios of the objects in $F_{12,phot} > 600 mJy$ regime.}
	\label{fig:600.pdf}
\end{figure}

1. For the stars in $F_{12,phot} > 600 mJy$ regime: 

There are 78 stars in total. We fit a normal distribution to the histogram of the flux density ratio $F_{excess}/F_{12,phot}$ as Fig.~\ref{fig:600.pdf} shows.
The resulting histogram has a standard deviation $\sigma =0.0354$, resulting in seven stars meeting Criterion B. 
Three of these had already been identified using criterion A, meaning that three new candidates are identified using criterion B. The bottom panel of Fig.~\ref{fig:figure2.pdf} compares the $3\sigma$ criterion B with the $3\sigma_{tot}/F_{12,phot}$ criterion A. This shows that the uncertainties given by $3\sigma_{tot}$ are a reasonably accurate measure of the true uncertainties in this regime, as measured empirically by $\sigma$, albeit that some stars, particularly the brightest ones (>10Jy) have a larger uncertainty, probably due to the stronger effects of saturation.

\begin{figure*}
	\includegraphics[width=0.9\textwidth]{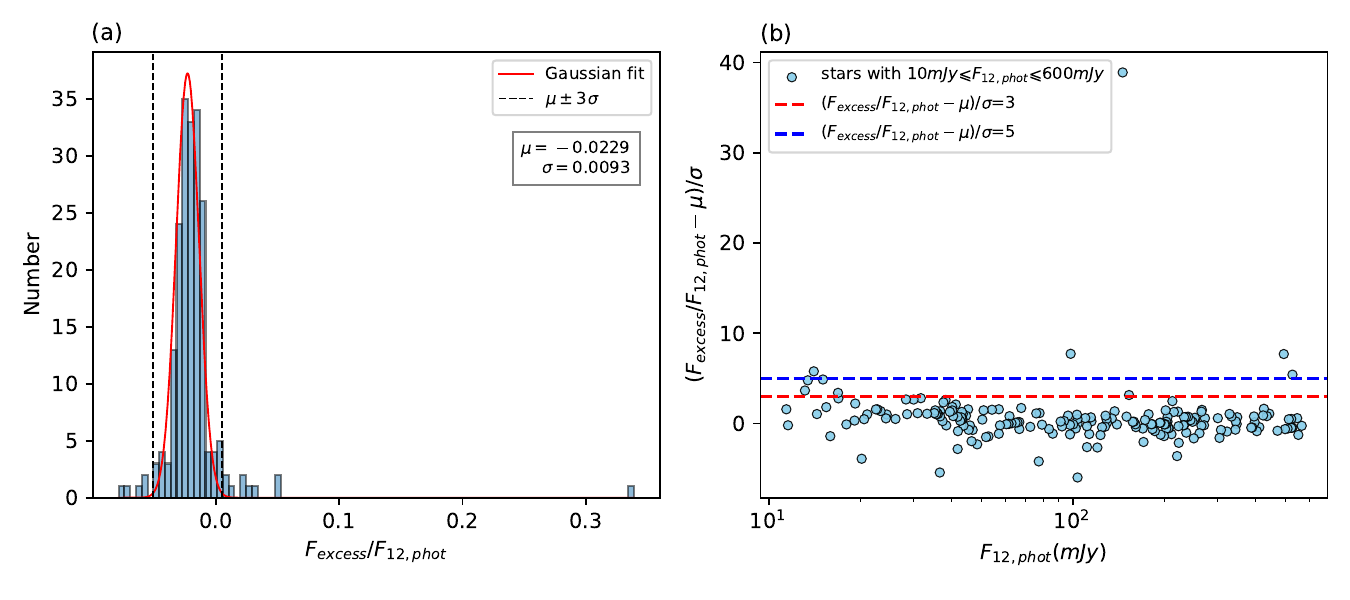}
	\caption{(a) Histogram of the flux density ratios of the objects in $10 mJy \leqslant F_{12,phot} \leqslant 600 mJy$ regime. (b) Scatter plot of $(F_{excess}/F_{12,phot} - \mu)/\sigma$ versus $F_{12,phot}$ for stars in $10 mJy \leqslant F_{12,phot} \leqslant 600 mJy$ regime. The dashed red and blue lines indicate the thresholds at $y=3$ and $y=5$, respectively.}
	\label{fig:10600.pdf}
\end{figure*}

2. For the stars in $10 mJy \leqslant F_{12,phot} \leqslant 600 mJy$ regime:

There are 203 stars in total. We fit a normal distribution to the histogram of the flux density ratio $F_{excess}/F_{12,phot}$ as Fig.~\ref{fig:10600.pdf} (a) shows. The resulting histogram has a standard deviation $\sigma = 0.0093$, resulting in ten stars meeting Criterion B. 

Given the dependence of the distribution of $F_{excess}/F_{12,phot}$ on $F_{12,phot}$ shown in Fig.~\ref{fig:figure2.pdf} for small and large $F_{12,phot}$, Fig.~\ref{fig:10600.pdf} (b) shows the same information zoomed in for the 10-600 mJy sample and compared with the mean and standard deviation of the distribution of the whole sample. This shows that the distribution is not significantly different across $F_{12,phot}$, although it is notable that more stars fall beyond $3\sigma$ at the extremes of $F_{12,phot}$ in this sample. 
We note that the distribution is wider in the 10–20 mJy regime, and thus Criterion B would underestimate the true $\sigma$. Therefore, we apply a more stringent cut of $B > 5$. In the 10–20 mJy regime, five stars meet Criterion B, but only one star (Scholtz's Star A) satisfies $B > 5$, so we remove the remaining four stars (LP 944-20, eps Ind B, 2MASS J00113182+5908400, UCAC4 379-100760).

Comparing the $3\sigma_{tot}/F_{12,phot}$ from Criterion A with the $3\sigma$ from Criterion B in Fig.~\ref{fig:figure2.pdf}, it is clear that in the $10 mJy \leqslant F_{12,phot} \leqslant 600 mJy$ regime, $3\sigma_{tot}/F_{12,phot}$ is approximately seven times the empirically determined uncertainties (i.e. $3\sigma$). We infer that this is because some of the calibration (or photosphere model) uncertainties have been overestimated. To assess how this might have affected the inferences of whether a star has an excess, we consider an alternative criterion A* which requires $\chi^{*}_{12}=F_{excess}/\sigma_{obs}>3$.

Criterion A only identifies one candidate in this regime, whereas Criterion A* identifies two new candidates. However, both of these new candidates had already been identified using Criterion B.

\begin{figure*}
	\includegraphics[width=0.9\textwidth]{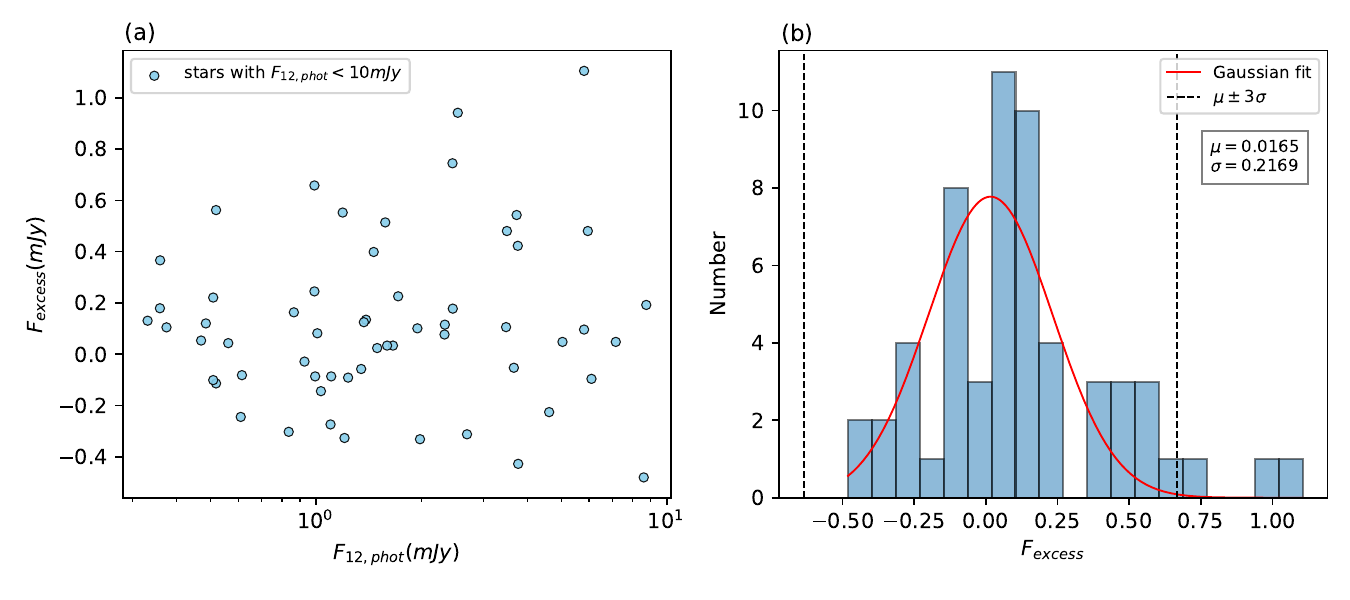}
	\caption{(a) Scatter plot of $F_{excess}$ versus $F_{12,phot}$ for stars in $F_{12,phot} < 10 mJy$ regime. (b) Histogram of the disc flux of the objects in $F_{12,phot} < 10 mJy$ regime.}
	\label{fig:10.pdf}
\end{figure*}

3. For the stars in $F_{12,phot} < 10 mJy$ regime:

There are 58 stars in total. However, Fig.~\ref{fig:figure2.pdf} shows that assuming $F_{excess}/F_{12,phot}$ is independent of $F_{12,phot}$ is going to return false positives. If this increase in uncertainty towards low fluxes is caused by the sensitivity limit of WISE, then the distribution of $F_{excess}$ should be uniform. This is tested and roughly confirmed in Fig.~\ref{fig:10.pdf} (a). We then fit a normal distribution to the histogram of the disc flux $F_{excess}$ as Fig.~\ref{fig:10.pdf} (b) shows. The resulting histogram has a standard deviation $\sigma_{excess} = 0.2169$, resulting in three stars that meet the criterion B. We note that, in order to be consistent with the other two regimes, we define the $\sigma = \sigma_{excess}/F_{12,phot}$ which is not a constant value (depending on $F_{12,phot}$) as Fig.~\ref{fig:figure2.pdf} shows. All of these were previously identified by criterion A, so no new candidates are identified in this way.

We also observe that for Criterion B, Fig.~\ref{fig:600.pdf} and Fig.~\ref{fig:10600.pdf} show many negative excesses below $5\sigma$. In Fig.~\ref{fig:600.pdf}, there are five such sources, and for the negative $5\sigma$ excesses, we attribute them to saturation effects, as this leads to flux values lower than the true flux. The W1 and W2 magnitudes of these five sources are all saturated (W1 < 8, W2 < 7). Therefore, we believe that including saturated sources results in a skewed distribution, as their fluxes are lower than predicted. 
But this reason does not apply to the interpretation of Fig.~\ref{fig:10600.pdf}, where both sources with negative excesses below $5\sigma$ are not saturated. As a result, we conducted statistical tests (e.g. Anderson-Darling and Shapiro-Wilk tests) to test the Gaussianity of the distributions in all regimes including Fig.~\ref{fig:600.pdf}, Fig.~\ref{fig:10600.pdf}, and Fig.~\ref{fig:10.pdf}, and found that none of them follows the Gaussian distribution. So it doesn’t make sense for there to be two negative excesses detections below $5\sigma$ in Fig.~\ref{fig:10600.pdf}.
This could mean that the significance of any positive excess above $3\sigma$ could be overestimated by assuming the distribution to be Gaussian, and we should remain cautious when interpreting excess identifications.

In summary, we list the 19 stars that meet Criterion A or Criterion B along with their parameters in Table~\ref{tab:criterion}. We exclud three stars that satisfied Criterion B but do not satisfy criterion A* within the regime of $10 mJy \leqslant F_{12,phot} \leqslant 600 mJy$. Then 16 stars that meet both Criterion A or Criterion B can be identified as promising candidates to explore further whether they exhibit excess emissions. We refer to these 16 stars as Sample C.

\begin{table*}
 \centering
 \caption{List of stars that meet Criterion A (or A*) or Criterion B.}
 \label{tab:criterion}
 \scalebox{0.9}
 {
  \begin{tabular}{lcccccccc} 
   \hline
   OBJ\_NAME & $F_{12,phot}$ &$F_{excess}$ & $F_{excess}/F_{12,phot}$ & $\sigma_{tot}/F_{12,phot}$ &  \multicolumn{3}{c}{Significance with Criterion} & Retain? \\
   \cline{6-8} 
    & $(mJy)$ & $(mJy)$ & &   & A & A*& B \\
   \hline
   $F_{12,phot} < 10 mJy :$&  &  & & &  \\
   \hline
   DENIS J025503.3-470049  &5.8120 &1.1048  &0.1901 &0.0614 &  3.10   &--- &5.02  & Yes \\
   2MASS J06073908+2429574  & 2.5362 &0.9415 &0.3712 &0.0931 & 3.99  &--- &4.26 &  Yes \\
   WISEP J180026.60+013453.1 &2.4497 &0.7445 &0.3039 &0.0998 &  3.04   &--- &3.36  &Yes \\
   WISEP J174124.25+255319.5  &0.9895 &0.6581 &0.6650 &0.1502 &  4.43    &---  &2.96  &Yes \\
   UGPS J052127.27+364048.6   &0.5188   &0.5620   &1.0833 &0.3046 &  3.56  & --- &2.51  &Yes \\  
   2MASSI J0340094-672405  &1.5755 &0.5140 &0.3262 &0.0847 & 3.85  &  ---  &2.29  &Yes \\
     \hline
   $10 mJy \leqslant F_{12,phot} \leqslant 600 mJy :$&  &  &  \\
   \hline
   CD-37 10765 B  &146.0210 &49.4742 &0.3388 &0.0641 & 5.29   &17.17 &38.89 & Yes \\
   Luhman 16 A &98.4681 &4.8033 &0.0488 &0.0495 &  0.99  &3.37  &7.71 &Yes \\
   mu. Her Ab  &494.7423 &24.0166 &0.0485 &0.0550 &  0.88   &3.14  &7.68  &Yes \\
   Scholtz's Star A  &14.0737 &0.4323 &0.0307 &0.0524 &  0.59  &1.35  &5.77  &No \\
   HD 103095  &528.3655 &14.4100 &0.0273 &0.0493 &0.55  &1.60  &5.39 & No \\
  SCR J1546-5534 B  &153.2260 &0.9607 &0.0063&0.049 &  0.13   &0.34  &3.14  & No \\
   \hline
   $F_{12,phot} > 600 mJy :$ &  &  &  \\
   \hline
   GJ 66 B &1959.7558 &1104.7643 &0.5637 &0.0711 &  7.93  & --- &16.82 & Yes \\
   70 Oph A  &4899.3894 &912.6314 &0.1863 &0.0562 &  3.32 & ---  &6.16 &Yes \\
   GJ 216 A  &3314.2929 &586.0054 &0.1768 &0.0560 &  3.16  & --- &5.89 &Yes \\
   41 Ara A  &1474.6202 &201.7396 &0.1368 &0.0524 &  2.61  & ---  &4.76&Yes \\
      HD 4628  &1234.4540 &168.1163 &0.1362 &0.0655 &  2.08 & --- &4.75& Yes \\
   HD 173739  &1340.1271 &177.723 &0.1326 &0.0527 &  2.52  & --- &4.64 & Yes\\
   36 Oph A  &3607.2956 &268.9861 &0.0746 &0.0492 &  1.52  & ---  &3.00 &Yes \\

   \hline
  \end{tabular}
 }
 
Notes. The ‘Retain?' column indicates whether the star is retained for further analysis based on the criteria met. Stars marked as ‘Yes' are retained for further study, while those marked as ‘No' are excluded.
\end{table*}

\section{Results: Vetting of exozodi candidates}
\label{sec:results}

In this section, we introduce additional criteria and filtering steps to further refine and validate our selection of excess emission candidates among the sources exhibiting IR excess. These procedures include WISE flags checking, visual inspections, positional offset analysis, examining the Figure of Merit (FoM) from Gaia and variability review, and help ensure that we focus on the most promising candidates.

\subsection{WISE flags checking}

We examined the WISE flags for source contamination/confusion (ccf) and extended source potential (ex), primarily to ensure the accuracy and reliability of the observational data. The ‘ccf' flag uses one character per band (W1/W2/W3/W4) to indicate whether the photometric and/or positional measurements of a source may be contaminated or biased due to proximity to image artifacts, with the digit ‘0' indicating that the source is unaffected by known artifacts. The ‘ex' flag represents the extended source flag, where a digit ‘0' signifies that the source shape is consistent with a point source and is not associated with or superimposed on a 2MASS XSC source. To ensure the greatest likelihood of the detected W3 excesses being legitimate, we retained only objects with a ‘0' in the ‘ccf' flag and a ‘0' in the ‘ex' flag, as shown in Table~\ref{tab:WISE flags}. This reduced the list of candidate objects to five.

\begin{table}
 \centering
 \caption{WISE flags for the 16 objects in Sample C.}
 \label{tab:WISE flags}
 \scalebox{0.9}
 {
  \begin{tabular}{lccc} 
   \hline
   OBJ\_NAME & ccf & ex  & Retain? \\
   \hline
   $F_{12,phot} < 10 mJy :$&  &  \\
   \hline
   DENIS J025503.3-470049 & 0000 & 0  & Yes \\
   2MASS J06073908+2429574 & 0000 & 0  &  Yes \\
    WISEP J180026.60+013453.1 & dd00 & 0 & Yes \\
   WISEP J174124.25+255319.5 & 0000 & 0   & Yes \\
   UGPS J052127.27+364048.6 & Hhh0 & 1   & No \\
   2MASSI J0340094-672405 & 0000 & 0   & Yes \\
   \hline
   $10 mJy \leqslant F_{12,phot} \leqslant 600 mJy :$& &  \\
   \hline
   CD-37 10765 B & 0000 & 1& No \\
   Luhman 16 A & 0h00 & 1 & No \\
   mu. Her Ab & dddd& 1 & No \\
   \hline
   $F_{12,phot} > 600 mJy :$ &  &  \\
   \hline
   GJ 66 B & hhd0 & 1& No \\
   70 Oph A & HHHH & 1  & No \\
   GJ 216 A & hhh0 & 1  & No \\
   41 Ara A & hh00 & 1  & No \\
   HD 4628& HHH0 & 0  & No \\
   HD 173739 & 0h00& 1  & No \\
   36 Oph A& hhh0 & 1  & No \\
   \hline
  \end{tabular}
 }
 
 Notes. The data, sourced from the ALLWISE catalogue, include the contamination/confusion (ccf) and extended source (ex) flags. Objects were either excluded or retained based on the values of these flags.
\end{table}

\subsection{Visual inspections}
\label{sec:Visual inspections}

A common technique for identifying and excluding unreliable sources is visual inspection of WISE images (e.g. \cite{2018ApJ...868...43S,2021MNRAS.508.3084S}), as not all flags provided by WISE can fully address issues in data reduction.

\cite{2024MNRAS.531..695S} categorized confounding sources into three types based on a detailed inspection of WISE images: blends, nebular features, and irregular structures. Figure 5 in \cite{2024MNRAS.531..695S} illustrates the characteristics and distinctions of each scenario, described as follows:

1. Blends: In the aperture of the WISE bands, particularly in the W3 and W4 bands, sources overlap with nearby sources. To identify these blended sources, high-resolution optical images can be used. Additionally, if an IR source is significantly offset from the image centre and lacks optical emission, it is considered a blended source, even if some contaminants do not emit optical light.

2. Nebular features: In this case, there is no discernible IR radiation source at the location of the candidate, and the W3 and W4 images typically appear hazy and disordered.

3. Irregular structures: Although selected based on WISE flags with ext\_flag values of 0, some images exhibit irregularities, particularly in the W3 and W4 bands, where the sources deviate from point-like shapes. These cases are classified as ‘Irregular'. In this category, the cause of irregularities in the W3 and W4 images is difficult to determine. For instance, large-scale images are examined to check for nebulosity, but no such features are detected in the surroundings. The exact cause of these irregularities remains unclear, and they are presumed to be related to factors such as high noise, faint nebular features, or blended sources.

\begin{figure*}
        \centering
	\includegraphics[width=0.9\textwidth]{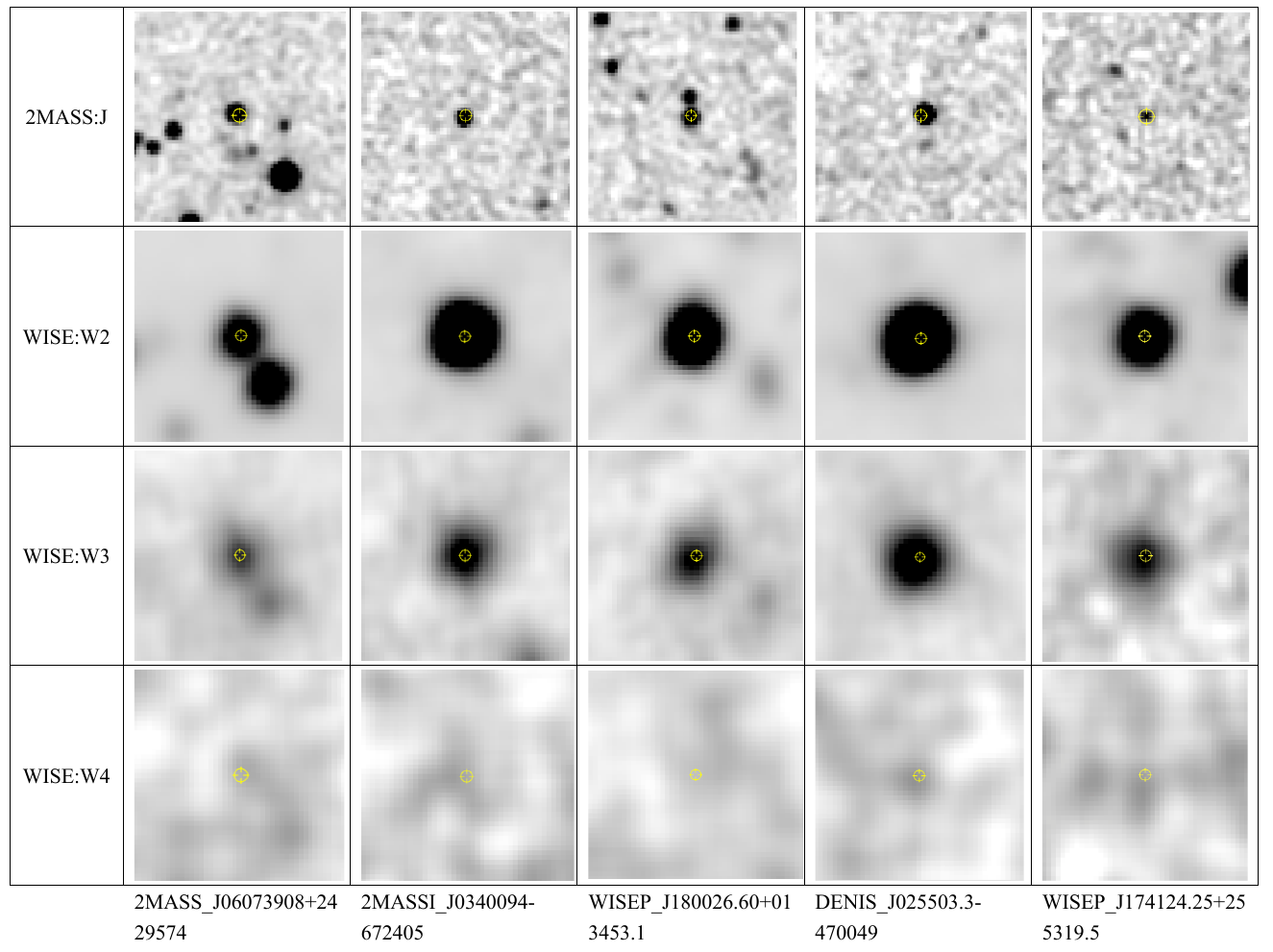}
	\caption{2MASS and WISE images of the five candidates. The size of all images is 1\arcmin. All images are centreed on the candidates' positions according to Gaia DR3. The central circle marks the location of the star as determined by Gaia DR3.}
	\label{fig:2 object}
\end{figure*}

Sources exhibiting any of the above conditions were excluded. After analysing the WISE images of the five candidates, none appeared to have conspicuous issues. In previous studies, numerous blend cases were identified through visual inspection of optical or NIR images \citep{2024MNRAS.531..695S}. Consequently, we performed a final visual inspection using the Aladin Sky Atlas \citep{2000A&AS..143...33B}. This software allows for easy viewing of images of the same source from various surveys. By examining optical and NIR images from Pan-STARRS1 \cite{2020ApJS..251....7F}, Sky Mapper \citep{2019PASA...36...33O}, PTF \citep{2011AJ....142...60V}, ZTF \citep{2020ApJS..249...18C}, SDSS \citep{2020ApJS..249....3A}, DSS2 \citep{2004A&A...426..367M}, and 2MASS \citep{2003yCat.2246....0C}, when available, we double-checked that the five candidates were free of contaminants. The results showed that the five objects do not exhibit obvious signs of contamination. Figure~\ref{fig:2 object} presents the 2MASS and WISE images of these five objects.

\subsection{Positional offset analysis}

Earlier studies have emphasized that Galactic background contamination and chance alignments with extragalactic sources can induce IR excess at the position of a star (e.g. \cite{2012MNRAS.426...91K}; \cite{2013ApJ...772...32K}). 
To evaluate Galactic background contamination, \cite{2013MNRAS.433.2334K} found that WISE source extractions within 2\degr\ of the Galactic plane were unreliable due to the high stellar density in this region of the sky. We checked the Galactic locations of the five objects and found that none were closer than 10\degr\ to the Galactic plane.

In addition to Galactic background contamination, chance alignments with sources that emit in the IR but are obscured in the optical represent another potential source of contamination. As mentioned in Sect.~\ref{sec:Visual inspections}, this can result in the observed ‘offset' effect. In order to assess the likelihood of such chance alignments, \cite{2012MNRAS.426...91K} compared galaxy counts with the number of their IR excess sources. 
\cite{2017AJ....153..165T} reanalysed the source extraction of their targets to determine the offsets between the W1, W2, and W3 images. These offsets were then compared to the inherent offsets of stationary objects, such as quasars, which serve as useful indicators of WISE’s astrometric precision due to their stationary position in the sky. The offset distribution of quasars closely resembled that of their disc candidate stars, both following a Gaussian distribution. One distribution represented the Right Ascension offset between the W1 and W3 positions ($\mu = 0\farcs08, \sigma= 5\farcs00$), while the other represented the Declination offset between the W1 and W3 positions ($\mu = -0\farcs21, \sigma = 5\farcs48$).

To estimate the probability of chance alignments with extragalactic sources, we adopted a similar method as \cite{2017AJ....153..165T} and conducted source extraction to determine the offsets among the W1, W2, and W3 images. 
For source position extraction, we followed the procedure outlined by \cite{2016ApJS..225...15C}. We downloaded 3 arcmin × 3 arcmin AllWISE W1–W3 images and used IRAF DAOFIND \citep{1986SPIE..627..733T,1993ASPC...52..173T} to extract the source position in each passband (using parameters fwhmpsf = 1, sigma = 0.1, threshold = 3.1, respectively). Table~\ref{tab:offset} summarizes the offsets between the positions of the extracted sources in different bands. It can be seen that for the (W1,W2) and (W1,W3) offsets, both in Right Ascension and Declination, the discrepancies are minimal and fall within the range obtained by \cite{2017AJ....153..165T}. 

This finding reduces the likelihood that these five objects are contaminated by chance alignments. However, given the small excess emission, it is reasonable to assume that most of the W1–W3 emission originates from the star itself and is well aligned. A chance alignment with a faint background object would likely produce a centroid shift smaller than the FWHM, which would still fall within the distribution described by \cite{2017AJ....153..165T}. Therefore, while we cannot entirely exclude the possibility of contamination from chance alignments, our analysis does not find evidence to suggest that such contamination is likely for these objects.

\begin{table}
	\centering
	\caption{Offsets in Right Ascension and Declination for the five candidates in the (W1,W2) and (W1,W3) bands, respectively.}
	\label{tab:offset}
	\scalebox{0.9}
	{
		\begin{tabular}{lccccc} 
			\hline
			OBJ\_NAME & (W1, W2) & (W1, W3) \\
			                     & offset (\arcsec) & offset (\arcsec) \\
			                     & RA, DEC & RA, DEC \\
			\hline
			2MASS J06073908+2429574 & 2.345, 0.347 & 1.859, 0.245 \\
			2MASSI J0340094-672405 & -0.084, 0.153 & 1.446, 0.767 \\
			WISEP J180026.60+013453.1 & -0.282, 0.293 & -2.179, 2.984 \\
			DENIS J025503.3-470049 & 3.399, 0.304 & 1.201, 2.182 \\
			WISEP J174124.25+255319.5 & -0.297, -0.877 & -0.438, 0.361 \\
			\hline
		\end{tabular}
	}
\end{table}

\subsection{Examining the FoM}

The next step involves examining the FoM provided by the ESA Gaia cross-match with AllWISE, following the recommendations by \cite{2020ApJ...891...97D}. The FoM is a dimensionless score generated by the Gaia cross-matching algorithm when associating Gaia sources with counterparts in other catalogues, such as AllWISE. This score quantitatively evaluates the likelihood or ‘quality' of a match, helping to indicate the reliability of cross-matched sources. A higher FoM suggests a more likely match, while a lower FoM indicates a less confident association. 

In the context of WISE data, this score is particularly useful for mitigating the issue of source confusion, where a source might appear aligned with a nearby unrelated object. This confusion can occur due to the larger point spread function (PSF) of WISE, making it challenging to distinguish between nearby sources, especially in crowded fields. \cite{2020ApJ...891...97D} suggest that a low FoM, particularly when accompanied by a high signal-to-noise ratio (SNR > 10), may be indicative of source confusion. As a rule of thumb, they recommend discarding cases where SNR > 10 and FoM < 4, as these are likely contaminated by nearby objects, causing an apparent infrared excess. 

\begin{figure}
	\includegraphics[width=0.5\textwidth]{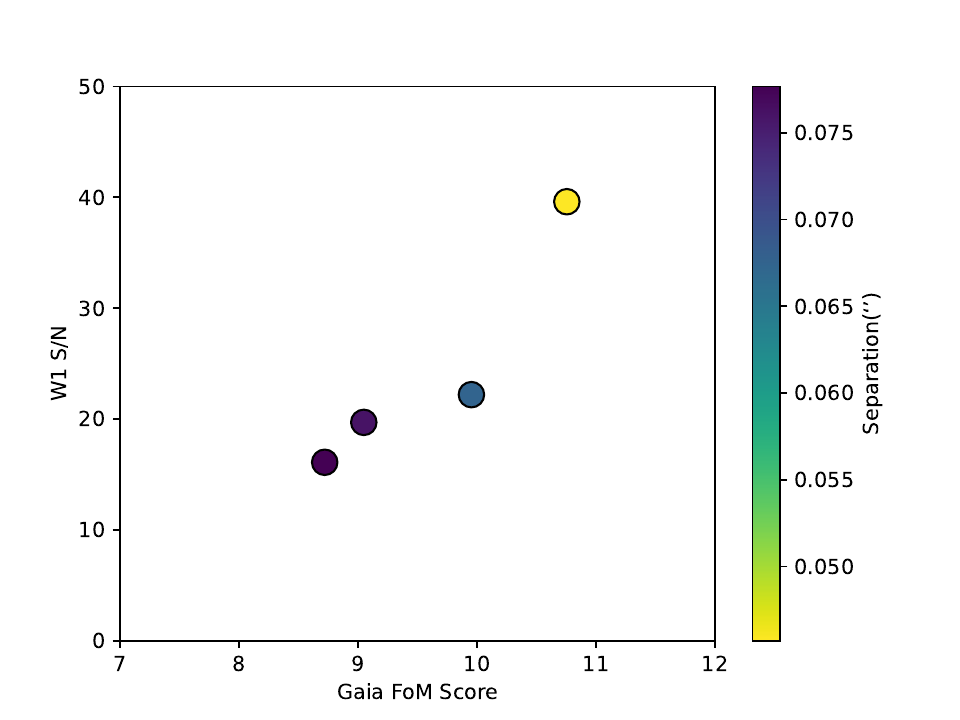}
	\caption{Gaia FoM for the five candidates (WISEP J174124.25+255319.5 does not have a FoM parameter) is plotted against the AllWISE S/N. The colour scale represents the separation between the expected position and the AllWISE detection. }
	\label{fig:FoM}
\end{figure}

As shown in Fig.~\ref{fig:FoM}, our analysis confirms that none of the five candidates in our sample have a FoM below this threshold. WISEP J174124.25+255319.5 is too faint to be detected by Gaia, so it does not have a FoM parameter. This indicates that source confusion is unlikely to be an issue for these candidates.

\subsection{Variability review}

The variability of WISE data can also lead to an apparent excess, so we also need to rule out this factor. We examined the variability characteristics using the variability flag (‘var') provided by WISE. The variability flag is a 4-character string, with each character corresponding to one band, representing the probability of flux variation based on individual WISE exposures. A value of ‘n' indicates insufficient or inadequate data to make a determination (fewer than six exposures). Values from 0 to 9 indicate increasing probabilities of variation: 0-5 represent sources that are most likely non-variable, 6-7 indicate possible variables (though susceptible to false-positive variability), and >7 indicate the highest probability of being true variables. The data show that none of the five candidates has a Variability flag >7 in the W3 band. 

We note that in the AllWISE Data Processing Explanatory Statement, there are some known issues with the WISE variability flag - however, most of these issues apply to assigning a potentially greater variability than is actually the case, rather than a lower variability which is what would cause a false positive exozodi detection. 
As a result, we conducted a basic variability search using WISE single exposure data. As shown in Appendix~\ref{sec:Variability and SED}, none of the five candidates exhibits obvious variability properties. Besides, we found the reported W3 flux value used in Sect.~\ref{sec:Identifying IR excesses} (Identifying IR excess) is consistent with the average of all measurements shown.

\subsection{Results}
\begin{figure}
        \centering
	\includegraphics[width=0.33\textwidth]{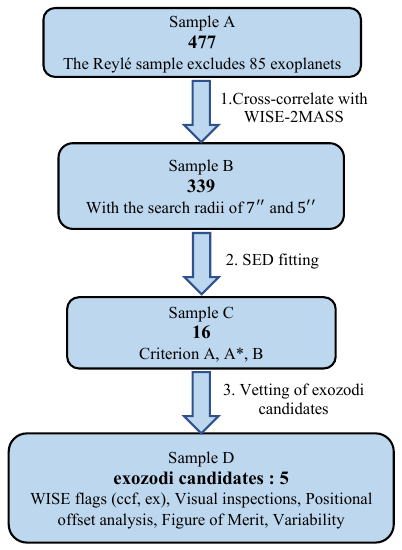}
	\caption{Flowchart highlighting our sample selection and vetting processes.}
	\label{fig:flowchart}
\end{figure}

In summary, our vetting process, has realized five objects being recognized as exozodi candidates, which we refer to as Sample D hereafter. Figure~\ref{fig:flowchart} illustrates the sample selection and vetting process we employed. The SEDs of these five exozodi candidates are shown in Appendix~\ref{sec:Variability and SED}. For the first three stars, the W3 value is significantly higher than the theoretical model, while this appears not to be the case for the last two stars. This discrepancy arises because the observed W3 flux (the red dot) should not be compared with the photospheric model prediction at that wavelength (the grey line), rather with the photospheric model that has been averaged over the W3 bandpass (the blue dot).

\section{Discussion}
\label{sec:Discussion}

\subsection{Exozodi candidate properties within Sample B}

\begin{figure*}
        \centering
	\includegraphics[width=1\textwidth]{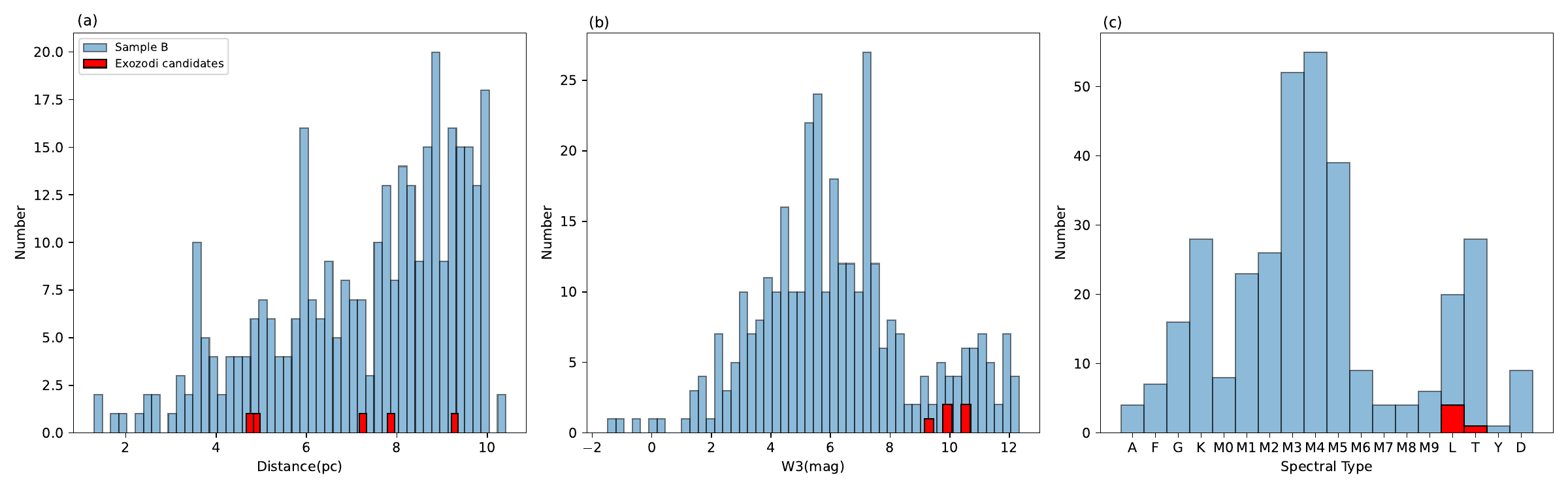}
	\caption{(a) Distribution of distances for the five exozodi candidates (in red) compared to those in Sample B (in sky blue). (b) Distribution of WISE W3 magnitudes for the five exozodi candidates (in red) relative to Sample B (in sky blue). (c) Distribution of spectral types for the five exozodi candidates (in red) in relation to Sample B (in sky blue).}
	\label{fig:Properties}
\end{figure*}

Figure~\ref{fig:Properties} shows the distribution of distances, WISE W3 magnitudes, and spectral types of the five exozodi candidates (Sample D) within its parent sample (Sample B). 
From panel (a), the distance distribution reveals that the five exozodi candidates are randomly distributed within 10 parsecs. Panel (b) shows the WISE W3 magnitudes distribution, which indicates that these candidates have W3 magnitudes concentrated around 10 mag, placing them on the fainter end of the parent sample, but these are not the faintest stars in that sample. Panel (c) shows the spectral type distribution, which reveals that these candidates are all BDs, belonging to a later type (but not the latest type) of the Sample B distribution. Overall, the most striking feature of these five exozodi candidates is their spectral type, all of which are BDs.

\begin{figure}
	\includegraphics[width=0.5\textwidth]{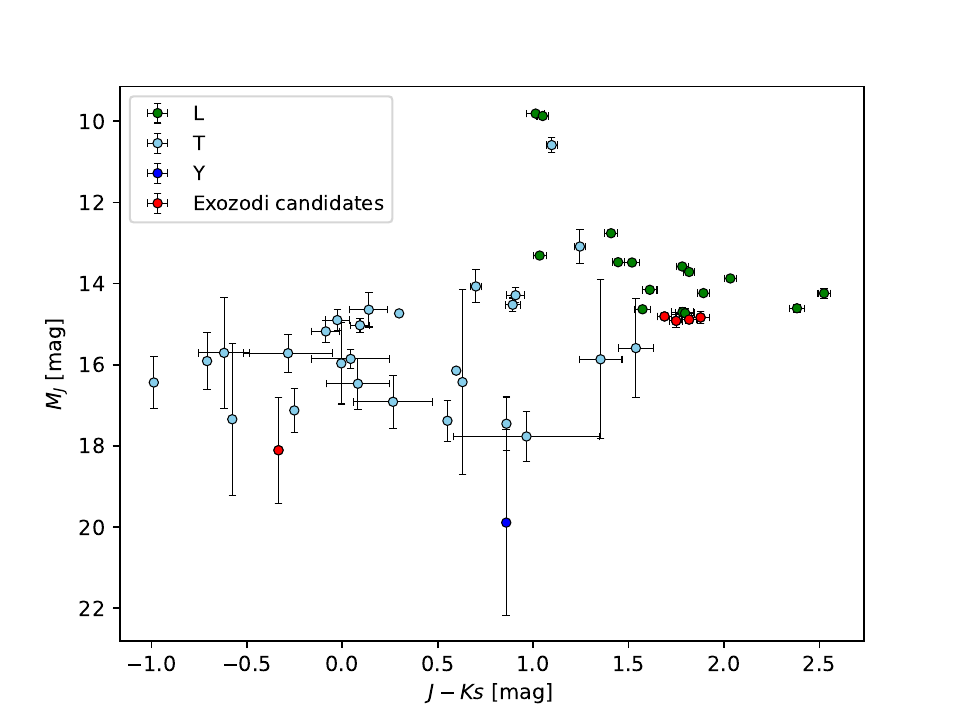}
	\caption{
Positions of the five exozodi candidates (in red) and the L, T, and Y-type BDs from Sample B (represented by green, sky blue, and blue, respectively) on the $J - K_s$ vs. $M_J$ colour-magnitude diagram.}
	\label{fig:jks}
\end{figure}

Figure~\ref{fig:jks} displays the distribution of the five exozodi candidates (in red) and the L, T, and Y-type BDs from Sample B (represented by green, sky blue, and blue, respectively) on a colour-magnitude diagram using 2MASS $J - K_s$ colour and absolute $M_J$ magnitude. Four of the exozodi candidates are concentrated in the range of $J - K_s$ colours from 1.5 to 2.0, and absolute magnitudes $M_J$ around 15, which is consistent with the typical location of L BD on this diagram. Another exozodi candidate is located at $J - K_s$ colour $< 0$ and an absolute magnitude $M_J$ around 18, which is the typical position for objects of spectral type $>T4.5$. As noted by \cite{2012ApJS..201...19D}, low-temperature BDs typically occupy this region on the $J - K_s$ vs. $M_J$ diagram, indicating that these candidates are cool and faint objects in the substellar regime typically associated with BDs.

For these five candidates, we collected the research content of previous publish papers as follows:

\textendash~2MASS J06073908+2429574: a late-type BD classified as L8 in optical and L9 in near-infrared spectra, located at a distance of 7.19 ± 0.10 pc \citep{2013ApJ...776..126C, 2016AJ....152..123G}. Its effective temperature is estimated to be 1250–1460 K, and its mass is below 0.055 $M_{\odot}$ \citep{2017ApJ...843..115R}. The presence of lithium absorption further supports an upper age limit of < 2 Gyr, indicating it is likely a young substellar object \citep{2013ApJ...776..126C, 2016AJ....152..123G}. The BD exhibits a bolometric luminosity of \(\log L/L_\odot = -4.66 \pm 0.02\) and a projected rotational velocity of \(v \sin i < 6 \, \text{km/s}\), suggesting it is likely viewed near its pole \citep{2016AJ....152..123G, 2017ApJ...843..115R}. Photometric monitoring by Spitzer and Kepler K2 detected no significant variability, with upper limits of \(<0.2\%\) and \(<1.4\%\), respectively \citep{2016AJ....152..123G}. Spectroscopic analysis revealed no \(\text{H}\alpha\) emission, with an equivalent width limit of \(<0.5 \, \text{\AA}\), indicating weak chromospheric activity \citep{2016AJ....152..123G, 2017ApJ...843..115R}. As an object in the L/T transition phase, it shows tentative \(\text{CH}_4\) absorption in the H band, which may provide insights into cloud dissipation and atmospheric evolution in ultracool BDs \citep{2013ApJ...776..126C}.

\textendash~2MASSI J0340094-672405: an L7-type BD whose distance was initially estimated as \(11.4 \pm 4.7\) pc based on parallax measurements from Spitzer and WISE data \citep{2007AJ....133..439C}. However, this estimate had a relatively large uncertainty. More recent Gaia DR3 \citep{2020yCat.1350....0G} parallax measurements refined the distance to \(9.358 \pm 0.038\) pc, confirming its membership in the 10 pc sample.

\textendash~WISEP J180026.60+013453.1: an L7.5$\pm$0.5 BD, classified based on both optical and near-infrared spectra. Parallax measurements place it at a distance of 8.01 $\pm$ 0.21 pc \citep{2015AJ....150..179G}. Lithium absorption (equivalent width of 2.7 Å) indicates a mass below 0.06 M$_\odot$, with an estimated age of 300--1500 Myr \citep{2015AJ....150..179G}. The effective temperature is 1430 $\pm$ 100 K, and its bolometric luminosity is log(L/L$_\odot$) $\approx$ -4.53 $\pm$ 0.04 \citep{2011AJ....142..171G}. Rotational measurements reveal a projected velocity of v sin i = 13.5 $\pm$ 0.5 km/s, corresponding to a rotation period of less than 9.3 hours \citep{2015AJ....150..179G}. Positioned in the L/T transition phase, this BD provides critical insights into the dynamics of substellar cloud evolution \citep{2015AJ....150..179G}.

\textendash~DENIS J025503.3-470049: a late-type L9 BD classified based on its spectral features, exhibits key characteristics such as CO and CH$_4$ absorption, along with shallow NH$_3$ and silicate cloud features characteristic of the L/T transition phase \citep{2004ApJS..154..418R, 2005ApJ...623.1115C}. Its estimated effective temperature is approximately 1400 K, with a high surface gravity of log g $\gtrsim$ 5, suggesting an age of 2--4 Gyr \citep{2016ApJ...817L..19T}. The projected rotational velocity is v sin i $\sim$ 40 km/s, indicating low angular momentum loss \citep{2014prpl.conf..433B}. High-resolution CRIRES+ K-band spectra reveal the presence of H$_2$O, CO, CH$_4$, and NH$_3$, with a $^{12}$CO/$^{13}$CO isotopologue ratio of 184$^{+61}_{-40}$, indicating a depletion of $^{13}$C relative to the local interstellar medium. Positioned at the L/T transition, it serves as a critical object for studying substellar atmospheric processes and cloud dynamics \citep{2024A&A...688A.116D}.

\textendash~WISEP J174124.25+255319.5: through near-infrared spectroscopy obtained with LBT/LUCIFER1, this object has been confirmed as a T9–T10 BD. Based on photometric estimates, its distance is approximately $4.6^{+1.2}_{-1.0}$ pc \citep{2011A&A...532L...5S}. The mass of this object is estimated to be in the range of 10–35 Jupiter masses ($M_{Jupiter}$). This target has been identified in both WISE and Spitzer near-infrared observations. However, observations in the LOFAR 111–169 MHz band did not detect this object, setting a $3\sigma$ upper limit of $<0.87 mJy$ \citep{2016MNRAS.463.2202B}.

\subsection{Accuracy of BD atmosphere predictions}

Our study indicates that exozodi candidates are very common around BDs, prompting discussions regarding the reliability of these detections. For example, it seems that 5/49 BDs in the Sample B have an excess. The low infrared flux of BD could lead to misidentifications, where background noise or contamination is more likely to be interpreted as excess emission. Through our vetting process in Sect.~\ref{sec:results}, we acknowledge that it is hard to confidently exclude contamination. However, since Fig.~\ref{fig:Properties} shows that these detections are not predominantly found towards the faintest IR sources, this suggests that contamination is unlikely to be a major issue. Alternatively, the phenomenon might be associated with inaccuracies in the photospheric flux predictions. Here we consider the accuracy of the BD atmospheric predictions.

Based on the BD atmosphere models available in VOSA, we evaluat the accuracy of the photospheric flux predictions. The study of BD atmospheres presents significant challenges due to the complexity of their low-temperature physics, chemical composition, and atmospheric processes. In this study, we utilize eight different atmospheric models to evaluate the SEDs of BDs in our sample. There are two main reasons for choosing these models: first, we were restricted to using models available in the VOSA database; second, through repeated experiments, we found that selecting these eight models together provides the most consistent fitting results for the BDs in our sample.

\cite{2022MNRAS.513.5701S} analysed the infrared spectra of several hundred M-type to T-type BDs observed with Spitzer. Their study revealed that silicate cloud absorption features are relatively common in L-type BDs. However, not all L-type BDs spectra exhibited these silicate absorption features.We have now entered the era of The James Webb Space Telescope (JWST), which provides new insights into BD atmospheres. The MIRI instrument aboard JWST allows for medium-resolution ($R\sim1500-3250$) mid-infrared (MIR) spectroscopic measurements \citep{2015PASP..127..584R}. Silicate absorption features have been detected with JWST MIRI, and \cite{2023ApJ...946L...6M} identified such features at 9 and 11 $\mu{m}$ in the BD VHS 1256 b.

\cite{2024ApJ...966L..11P} conducted an analysis of VHS 1256b data, employing five different forward models: ATMO, Exo-REM \citep{2018ApJ...854..172C}, Sonora Diamondback \citep{2024ApJ...975...59M}, BT-Settl, and DRIFT-PHOENIX. None of these models were able to reproduce the silicate cloud absorption features observed in VHS 1256b. \cite{2025A&A...694A.275C} introduced a new set of MSG models aimed at addressing this issue. These models have shown the ability to approximate the influence of clouds on BD atmospheres in the latest JWST observations, although they still do not fully reproduce the silicate features.

As a result, in our VOSA SED fitting, silicate cloud features are not fully reproduced, which may increase the uncertainty of the BD model prediction flux from such sources. 
Among the five exozodi candidates, silicate cloud features have previously been detected in DENIS J025503.3-470049 \citep{2004ApJS..154..418R, 2005ApJ...623.1115C}, suggesting the W3 excess may not be true. For the remaining four sources, no silicate signatures have yet been reported, so we cannot determine whether the W3 band excess arises from exozodi emission or from silicate clouds. However, since these excesses all arise towards stars in a specific spectral type range, for which atmospheric anomalies at this wavelength are known, we conclude that it is most likely that these are not true exozodi candidates. Thus, we interpret the WISE observations of these five objects as providing upper limits to the exozodi flux levels. And in what follows, we consider these as non-detections of exozodi, although these candidates merit spectroscopic follow-up to confirm this interpretation. We also note that improved atmosphere models are needed to make the accurate photospheric flux predictions required to search for the presence of faint exozodi levels and suggest that a more stringent $\chi_{12} >5$ criterion should be used for excess identification for this spectral type range (LTY).

\subsection{Exozodi detection rate}

In this section, we compare the upper limit from our Sample A with the detection rate from KW13 sample. It is important to recognize the main differences between the samples used in each analysis when doing this comparison. Both samples are volume-limited but KW13 analysed only solar-type stars within 120 parsecs, while our Sample A which belongs to the Reyl\'{e} sample includes nearby stars across nearly all spectral types within 10 parsecs (see Fig.~\ref{fig:Properties}). Indeed, 52\% of stars in Sample A, are M-type stars. 
Also, the KW13 sample consists of 24,174 stars, whereas Sample A contains only 477 stars.

The lack of exozodi detections in our sample B, with zero detections out of 339 stars, is consistent with the detection rate in the KW13 sample, which is 25 detections out of 24,174 stars ($\sim10^{-3}$). To quantify this, we calculate the probability of obtaining no detections in a sample of 339 stars, given a detection rate $10^{-3}$, which is $(1-10^{-3})^{339}=71\%$. This probability indicates that it would have been likely to get no detection if the rate of exozodi was the same for our sample as for the KW13 sample.

However, it is important to note that Kennedy \& Wyatt did not measure the luminosity function for low-mass M-type stars, which dominate our sample. Therefore, our results provide the first constraints on the exozodi luminosity function for such stars. In the absence of evidence suggesting that the luminosity function for low-mass stars differs from that of Sun-like stars, we conclude that the best estimate is that the exozodi luminosity function is independent of spectral type, although what we can say for sure is that <0.0044(1:226) M stars have 12 $\mu{m}$ excesses in exozodi levels > 0.21 times the stellar photosphere (the mean of upper limits in Table~\ref{tab:339 information} for M stars).

\subsection{Implications of exozodi limits on searches for exo-Earths}

\begin{figure}
	\includegraphics[width=0.48\textwidth]{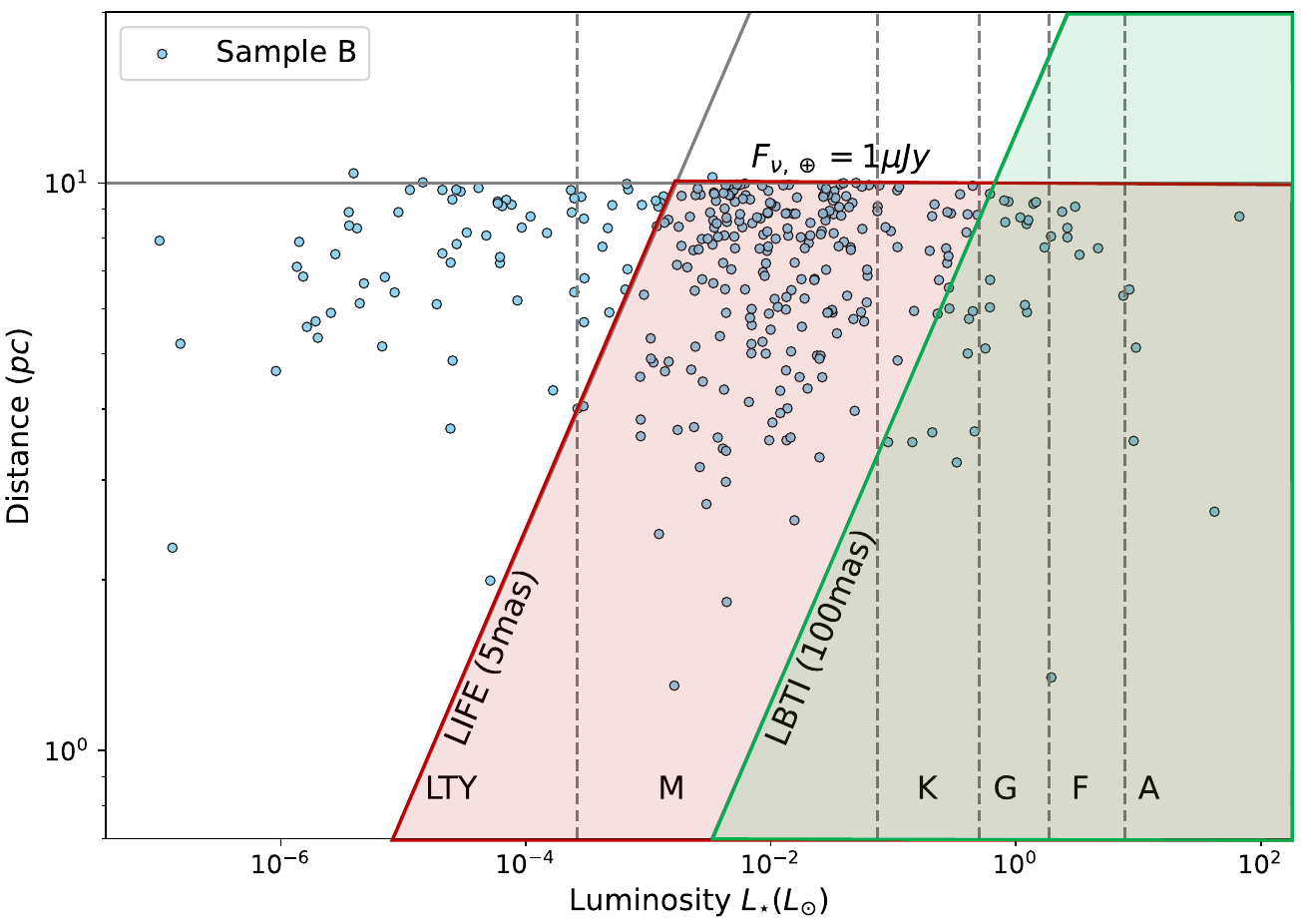}
	\caption{Distance versus stellar luminosity plot for Sample B. Skyblue dots represent the entire Sample B. The two diagonal lines represent the detection limits of the LIFE, and LBTI, respectively. The horizontal line at a distance of 10 parsecs corresponds to the distance at which an exo-Earth would emit a flux density of 1 $\mu{Jy}$ at a wavelength of 12 $\mu{m}$.}
	\label{fig:location}
\end{figure}

Imaging exo-Earths is primarily limited to stars located relatively close to Earth, so our Sample B is the ‘super-sample' from which many targets are likely to be drawn. This sample is summarized in Fig.~\ref{fig:location}, which shows the distribution of stars in terms of their luminosity and distance. This figure can be used to assess the detectability of exo-Earths, since a simple blackbody approximation can be used to determine how the angular scale of the HZ of a star depends on its luminosity and its distance from the solar system. Treating both an exo-Earth and its host star as blackbodies, the orbital distance (representing the HZ radius) of a $\sim300K$ exo-Earth can be estimated by equating the proportion of the star’s luminosity received by the planet to its own blackbody emission,

\begin{equation}
	\frac{a_{HZ}}{au}=\sqrt{\frac{L_{\star}}{L_{\sun}}},
	\label{equation:2}
\end{equation}

\noindent with the angular size at that HZ given by

\begin{equation}
	(\theta_{HZ}/\arcsec)=\sqrt{{L_{\star}}/{L_{\sun}}} /(d/pc).
	\label{equation:3}
\end{equation}

This means that for a given angular resolution of a particular telescope, there is a diagonal line in the $\log(d)$-$\log(L)$ plane that describes for which stars this telescope would be able to resolve its HZ. For example, the LIFE nulling interferometer has a maximum angular resolution of 5 mas \citep{2022A&A...664A..21Q} as illustrated with the diagonal red line in Fig.~\ref{fig:location}. It is only below such lines that warm dust and exo-Earths can be spatially resolved from their host star with this telescope, while above them, their emission lies too close to the central star and so detection is not possible. The horizontal line at a distance of 10 parsecs corresponds to the distance at which an exo-Earth would emit a flux density of 1 $\mu{Jy}$ at a wavelength of 12 $\mu{m}$. This represents a further practical consideration, due to the integration time required, if one were to detect the 12 $\mu{m}$ emission from an exo-Earth; that is exo-Earths could be imaged in the infrared at larger distances but might require prohibitive integration times. Thus the targets for which LIFE might be able to resolve and image a habitable exo-Earth lie in the region of Fig.~\ref{fig:location} shown with the red shading.

For future exo-Earth imaging missions, it is crucial to determine whether the stars they would be searching host exozodis. Our analysis indicates that none of the LIFE candidates show clear evidence of exozodis in WISE data. However, this does not rule out the presence of exozodis. For instance, the Large Binocular 
Telescope Interferometer (LBTI) \citep{2016SPIE.9907E..04H} used nulling interferometry to detect fainter excess levels than the $\sim10\%$ excess we were able to detect with WISE photometry; for example, \cite{2020AJ....159..177E} found that 20\% of stars exhibit excess levels as low as 0.1–1\%. However because this technique relies on resolving the emission, the 100 mas resolution of LBTI means that it was only able to search for exozodis within the green-shaded region; that is towards spectral types earlier than K. Thus, the LBTI cannot be used to search for faint exozodi for all potential exo-Earth imaging targets, notably the HZs of M-type stars and later types cannot be probed by LBTI.

Figure~\ref{fig:location} shows that it will not be possible to resolve the emission from an exo-Earth in the HZ of BD even with LIFE. As such, the existence (or not) of exozodiacal dust around such stars is not yet a significant concern. This figure also indicates that the majority of targets for which LIFE will be capable of resolving exo-Earths are M-type stars. Until now, no measurements of the exozodi luminosity function for such stars have been made, although searches for bright discs around M stars out to large distances have been performed \citep{2021MNRAS.508.3084S}. In Table~\ref{tab:339 information}, we present the first upper limits on exozodi emission towards all stars in the Sample B, for which we use the larger value between $3\sigma_{tot}/F_{12,phot}$ and $3\sigma$ as the upper limit on $F_{excess}/F_{12,phot}$ at 12 $\mu{m}$.

We find that less than 1\% of M-type stars have exozodi emission exceeding approximately 21\% excess levels, which is consistent with the rate of exozodi at similar levels for FGK stars. Give the inability of LBTI to probe later spectral types, Fig.~\ref{fig:location} shows that we currently have no constraints on faint exozodi levels around M-type stars, nor any prospects for obtaining such constraints. Nevertheless, one promising way to improve these limits would be through a mid-infrared spectroscopic survey using the JWST.

\section{Conclusion}
\label{sec:Conclusion}

In this study, we conducted a thorough search for exozodi candidates within 10 parsecs, starting with the Reyl\'{e} sample, which spans most spectral types, and performing cross-matching with the WISE-2MASS-Gaia database. Through various analytical methods, we identified five promising upper limit exozodi. Despite the limitations of current atmospheric models for BDs, we found that the silicate cloud features observed in DENIS J025503.3-470049 likely indicate that its inferred W3 excess is not because of its exozodi emission but the underestimated BD model uncertainty. Future observations will be required to confirm whether other candidates, such as 2MASS J06073908+2429574, 2MASS J0340094-672405, WISEP J174124.25+255319.5, and WISEP J180026.60+013453.1, exhibit true exozodi signatures. However, given the stringent $5\sigma$ criterion needed for a confident exozodi excess, we concluded that none of the objects in our sample show such excesses.

Furthermore, less than 1\% of M stars have exozodi emissions above 21\% excess levels, which is consistent with the rate of exozodi at similar level towards FGK stars in KW13 sample, and no significant differences in the distribution of lower-mass stars compared to higher-mass stars were observed. The KW13 sample did not measure the luminosity function for low-mass M stars, which happen to be the dominant type in our sample. 
Therefore, our results provide the first constraints on the exozodi luminosity function for such stars. In the absence of evidence for this being different from that of Sun-like stars we conclude that the best estimate is for the luminosity function to be independent of spectral type.

For future exo-Earth imaging missions, LIFE will be capable of resolving exo-Earths around most M-type stars but will not be able to resolve the emission from exo-Earths in the HZ of BDs. Exozodis can be detected at much lower levels using nulling interferometric techniques with LBTI, but its spatial resolution is only sufficient for FGK-type stars. Therefore, we have no constraints on faint exozodi levels around M-type stars and no prospects for obtaining such constraints. However, one way to improve these limits with currently available instrumentation is through mid-infrared spectroscopic surveys with JWST. The potential of future observations with Extremely Large Telescope (ELT)/METIS or Very Large Telescope Interferometer (VLTI)/NOTT should also be considered. We provide the first upper limits on exozodi emissions for all stars in Sample B (the sample of 339 objects identified through proper motion calculations and cross-matching with the WISE and 2MASS databases), offering important constraints for future studies of exozodi. Without exozodi detections, it was not possible to address the goal of identifying conditions that correlate with exozodi dust levels, however, the results of this study lay a solid foundation for further investigation into the characteristics and distribution of exozodi around nearby stars.

\section{Data availability}
\label{sec:Data availability}
Table~\ref{tab:339 information} is only available in electronic form at the CDS via anonymous ftp to cdsarc.u-strasbg.fr (130.79.128.5) or via \url{http://cdsweb.u-strasbg.fr/cgi-bin/qcat?J/A+A/.}

\begin{acknowledgements}

The authors thank the referee for careful reading of the manuscript and feedback, which improved the clarity of this paper. We also like to thank Paul Storrs for initial work on the project which provided the motivation for this work. This work was supported in part by the NSFC under grant U1631109 and in part by Guizhou Provincial Basic Research Program (Natural Science) (No.MS[2025]694) and Guizhou Provincial Science and Technology Projects (No.QKHFQ[2023]003 and No.QKHPTRC-ZDSYS[2023]003).

This publication makes use of VOSA, developed under the Spanish Virtual Observatory (\url{https://svo.cab.inta-csic.es}) project funded by MCIN/AEI/10.13039/501100011033/ through grant PID2020-112949GB-I00. VOSA has been partially updated by using funding from the European Union's Horizon 2020 Research and Innovation Programme, under Grant Agreement nº 776403 (EXOPLANETS-A).

This research has made use of the NASA/IPAC Infrared Science Archive, which is funded by the National Aeronautics and Space Administration and operated by the California Institute of Technology.

This research has made use of the VizieR catalogue access tool, CDS, Strasbourg, France \citep{10.26093/cds/vizier}. The original description of the VizieR service was published in \citet{vizier2000}.

This work presents results from the European Space Agency (ESA) space mission Gaia. Gaia data are being processed by the Gaia Data Processing and Analysis Consortium (DPAC). Funding for the DPAC is provided by national institutions, in particular the institutions participating in the Gaia MultiLateral Agreement (MLA). The Gaia mission website is \url{https://www.cosmos.esa.int/gaia}. The Gaia archive website is \url{https://archives.esac.esa.int/gaia}.

This research has made use of ‘Aladin sky atlas' developed at CDS, Strasbourg Observatory, France \cite{2000A&AS..143...33B} (Aladin Desktop).

This research use the \tt{PYTHON} packages \texttt{ASTROPY} and \texttt{ASTROQUREY} \citep{2012ascl.soft08017R}, \texttt{MATPLOTLIB} \citep{2007CSE.....9...90H}, \texttt{PANDAS} \citep{pandas}, \texttt{NUMPY} and \texttt{SCIPY} \citep{5725236}.

\end{acknowledgements}

\bibliographystyle{aa}
\bibliography{references}
\onecolumn
\begin{appendix}
\section{Light curves and SEDs of the five exozodi candidates.}
\label{sec:Variability and SED}

\begin{figure*}[!ht]
	\includegraphics[width=1\textwidth]{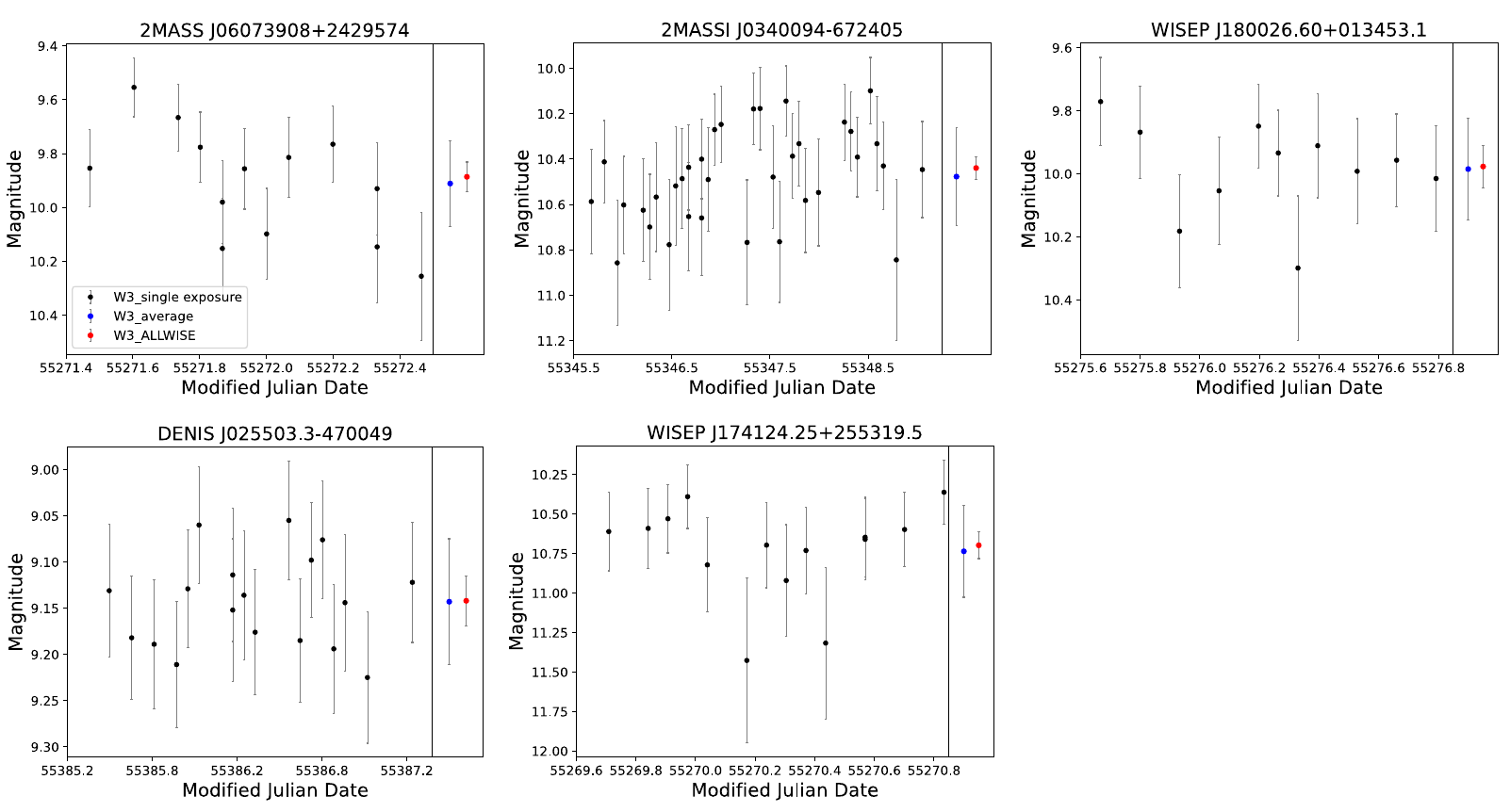}
	\caption{Light curves of the five targets observed with the WISE W3 band. Each panel shows the magnitude variation over Modified Julian Date. Error bars represent measurement uncertainties. We also plot the W3 flux value (and its associated error) used in Sect.~\ref{sec:Identifying IR excesses} (Identifying IR excess) directly in the right region of each panel as the red dot shows, as well as the blue dot which represents the average of all single exposure measurements data for comparison.}
	\label{fig:Variability}
\end{figure*}

\begin{figure*}[!ht]
	\includegraphics[width=1\textwidth]{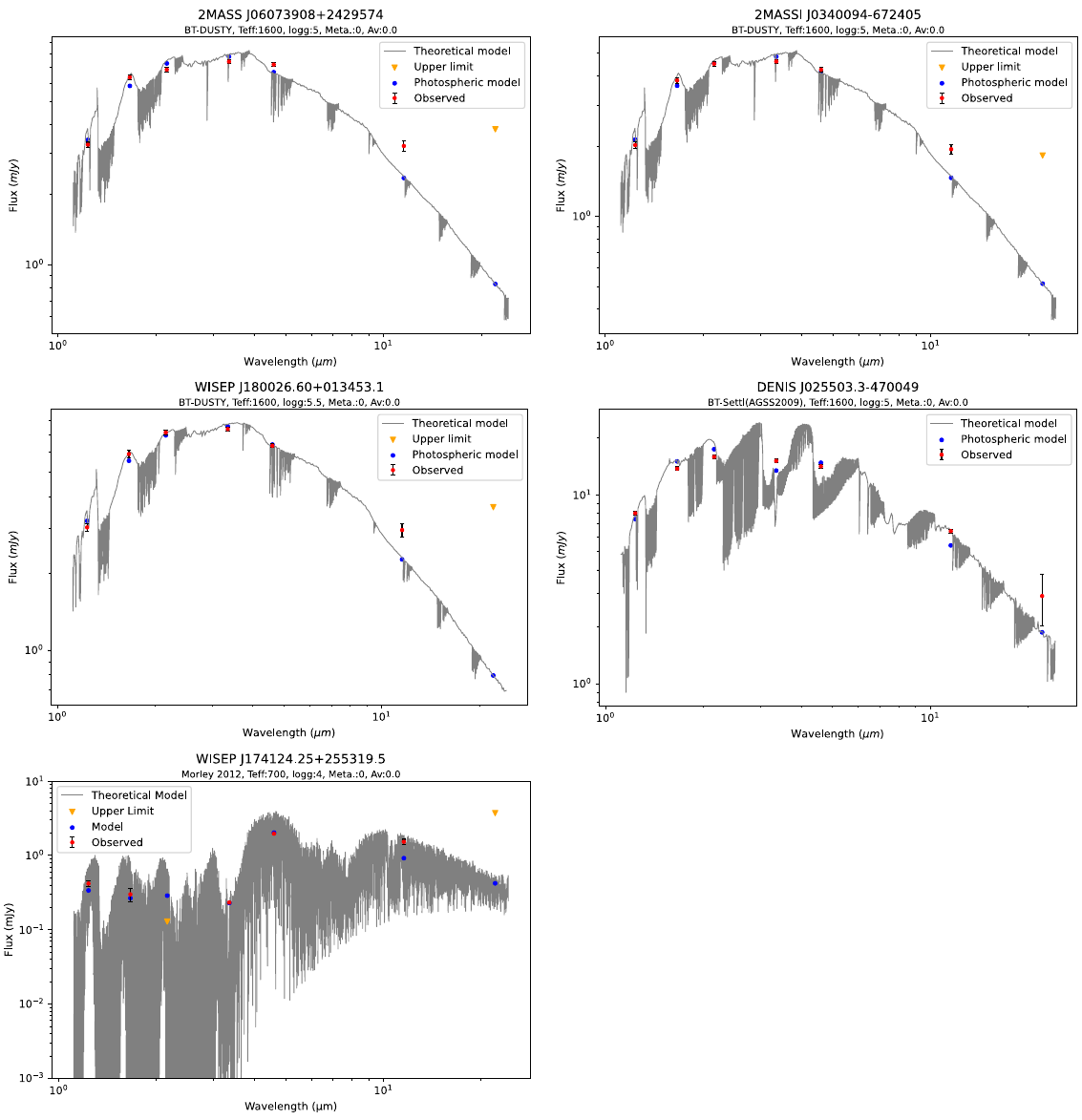}
	\caption{SEDs for the five exozodi candidates. The red dots represent the observed photometry in the NIR and IR bands ($J$,$H$,$K_s$, $W1$,$W2$,$W3$ and $W4$). The inverted yellow triangles indicate upper limits, which are not used for the fitting. The blue dots represent the best-fit photospheric model flux, with the theoretical spectrum plotted behind it in gray.}
	\label{fig:SED}
\end{figure*}

\end{appendix}

\end{document}